\documentclass[twocolumn]{aastex631}

\usepackage{multirow}

\usepackage[utf8]{inputenc}
\usepackage{natbib}
\usepackage{graphicx}
\usepackage{longtable}
\usepackage{enumerate}
\usepackage{amsmath}
\usepackage{hyperref}
\usepackage{CJK}
\usepackage[title]{appendix}
\usepackage{rotating}
\usepackage{longtable}

\usepackage{amsmath}

\def\msun{\ifmmode {\rm M}_{\mathord\odot}\else $M_{\mathord\odot}$\fi}
\def\rsun{\ifmmode {\rm R}_{\mathord\odot}\else $R_{\mathord\odot}$\fi}
\def\lsun{\ifmmode {\rm L}_{\mathord\odot}\else $L_{\mathord\odot}$\fi}

\def\c18o{C$^{18}$O}
\def\h2{H$_{2}$}
\def\13co{$^{13}$CO}

\def\radmc{{\sc radmc-3d}}
\def\cm2{cm$^{-2}$}
\def\cmc{cm$^{-3}$}

\newcommand{\um}{$\mu$m}

\newcommand{\hii}{H\,{\sc ii}}

\def\um{$\mu$m}
\newcommand{\CASI}{{\sc casi}}

\def\deg{$^{\circ}$}

\shorttitle{}
\shortauthors{}

\begin{document}
\begin{CJK*}{UTF8}{gbsn}

\title{Predicting the Radiation Field of Molecular Clouds using Denoising Diffusion Probabilistic Models}

\author[0000-0001-6216-8931]{Duo Xu}
\affiliation{Department of Astronomy, University of Virginia, Charlottesville, VA 22904-4235, USA}

\author[0000-0003-1252-9916]{Stella S. R. Offner}
\affiliation{Department of Astronomy, The University of Texas at Austin, Austin, TX 78712, USA}

\author[0000-0002-6447-899X]{Robert Gutermuth}
\affiliation{Department of Astronomy, University of Massachusetts, Amherst, MA 01003, USA }

\author[0000-0002-1655-5604]{Michael Y. Grudi\'{c}}
\affiliation{Carnegie Observatories, 813 Santa Barbara St, Pasadena, CA 91101, USA}

\author[0000-0001-5541-3150]{D\'{a}vid Guszejnov}
\affiliation{Center for Astrophysics | Harvard \& Smithsonian, 60 Garden St, Cambridge, MA 02138, USA}

\author[0000-0003-3729-1684]{Philip F. Hopkins}
\affiliation{TAPIR, Mailcode 350-17, California Institute of Technology, Pasadena, CA 91125, USA}

\email{xuduo117@virginia.edu}

\begin{abstract}

Accurately quantifying the impact of radiation feedback in star formation is challenging. To address this complex problem, we employ deep learning techniques, denoising diffusion probabilistic models (DDPMs), to predict the interstellar radiation field (ISRF) strength based on three-band dust emission at 4.5 \um, 24 \um, and 250 \um. We adopt magnetohydrodynamic simulations from the STARFORGE (STAR FORmation in Gaseous Environments) project that model star formation and giant molecular cloud (GMC) evolution. We generate synthetic dust emission maps matching observed spectral energy distributions in the Monoceros R2 (MonR2) GMC. We train DDPMs to estimate the ISRF using synthetic three-band dust emission. The dispersion between the predictions and true values is within a factor of 0.1 for the test set. {We extended our assessment of the diffusion model to include new simulations with varying physical parameters. While there is a consistent offset observed in these out-of-distribution simulations, the model effectively constrains the relative intensity to within a factor of 2.} Meanwhile, our analysis reveals weak correlation between the ISRF solely derived from dust temperature and the actual ISRF. We apply our trained model to predict the ISRF in MonR2, revealing a correspondence between intense ISRF, bright sources, and high dust emission, confirming the model's ability to capture ISRF variations. {Our model robustly predicts radiation feedback distribution, even in complex, poorly constrained ISRF environments like those influenced by nearby star clusters. However,  precise ISRF predictions require an accurate training dataset mirroring the target molecular cloud's unique physical conditions.}

%%{another more big-picture type thing for us to think about: I assume there are more classical approaches to mapping dust emission to radiation field intensity - can we compare the performance with one of these models?}{SO: One could assume the dust is a GB or BB but that provides a poor solution since its not when you have resolved embedded stellar sources. E.g., Barnard ea 2010 provides a simple mapping between TDust and the ISRF but their focus is the long-wavelength portion of the SED, in which the influence of the massive stars is indirect.  This long wavelength emission is usually used to infer temperature but this is again not necessarily well-correlated with the actual radiation field strength either (as this exercise shows in Fig. 11).}

\end{abstract}


\section{Introduction}
\label{Introduction}

Stellar feedback plays a crucial role in the star formation process, manifesting in two main forms: mechanical feedback and radiative feedback \citep{2010ApJ...710L.142F,2020SSRv..216...68G}. Mechanical feedback involves the injection of momentum and kinetic energy into the surrounding clouds through stellar winds, including protostellar outflows and isotropic stellar-wind-driven bubbles \citep{2007ApJ...670..428C,2007prpl.conf..245A,2014prpl.conf..451F}. Conversely, radiative feedback is associated with the dissociation and ionization of cold molecular gas by the intense radiation emitted by massive stars \citep{2012MNRAS.427..625W,2019MNRAS.488.2970G,2020AJ....160...78R}. This radiation also exerts pressure on the surrounding gas and dust, resulting in the formation of ionized bubbles known as \hii\ regions, which release a substantial amount of energy \citep{2014ApJ...795..121L}.

Recent studies have highlighted the significant impact of stellar feedback on the star formation process. Simulations show that mechanical feedback, such as outflows and stellar winds, reduces protostellar masses and accretion rates, as well as disperses surrounding gas, leading to a decrease in both the global star formation rate and efficiency \citep{2007ApJ...659.1394M,2014ApJ...790..128F,2015MNRAS.450.4035F,2017ApJ...847..104O,2022MNRAS.515.4929G}. However, the energy injection from outflows is typically limited to smaller scales, ranging from sub-parsec to parsec scales \citep{2010ApJ...709...27W,2022ApJ...926...19X}. In contrast, the combined effects of photoionization and radiation pressure from massive stars and their \hii\ regions result in the heating of the surrounding gas and efficient dispersal of the nearby cloud \citep{2012MNRAS.424..377D,2013MNRAS.430..234D}. The radiation feedback from massive stars can have a broad impact, spanning scales from a few parsecs to tens of parsecs \citep{2012MNRAS.427..625W,2014ApJ...795..121L,2020AJ....160...78R,2022MNRAS.515.4929G,2022MNRAS.512..216G,2022ApJ...941..202R}. %%It is important to mention that the expansion of \hii\ regions has been suggested to potentially have a positive impact by triggering new star formation \citep{2006MNRAS.367..763S}. However, the extent of this influence is still a subject of debate, with some suggesting that it may be minimal \citep{2007MNRAS.375.1291D}.

In order to gain a comprehensive understanding of how stellar feedback influences the star formation process, including star formation rate and efficiency, it is crucial to study feedback mechanisms across molecular clouds with varying physical and chemical conditions. However, accurately quantifying the impact of radiation from massive stars continues to present a challenge in observational studies. There are several current ``classical" approaches to estimating the radiation field from observations. For example, the strength of the radiation field originating from massive stars is commonly estimated using dust emission \citep{2010A&A...518L..88B}. However, the mean dust temperature as derived from long-wavelength emission gives an incomplete picture of local conditions. \citet{2023AJ....165...25P} developed a framework using the ratio between far-infrared (FIR) fine-structure lines, such as [O I], [C I], and [C II], to estimate the strength of the radiation field through photodissociation region (PDR) models. However, PDR models rely on simplified assumptions about the cloud geometry and density distribution, leading to uncertainties when applied to actual observational data. This approach also does not provide an accurate estimate of the radiation field within the cloud due to young embedded sources. Other PDR codes, such as 3D-PDR \citep{2012MNRAS.427.2100B}, offer the advantage of allowing for arbitrary density distributions and the ability to specify radiating sources within the cloud, addressing some of the uncertainties mentioned earlier. However, degeneracy due to physical conditions imposes a significant limitation in using line ratios to determine the strength of the radiation field, since different number densities and radiation field strengths can produce the same line ratio \citep{2023AJ....165...25P}. Additionally, mapping FIR lines across molecular clouds is time-consuming, especially in quiescent regions where [O I], [C I], and [C II] emission is relatively faint. 
%{mike: the discussion goes from machine learning to classical approaches then back to ML - maybe mention the classical stuff first and then introduce ML and proceed from there to outline our study. DX: I switched the order.}

%SO I put this back
By comparison, machine learning provides a promising avenue for improving the estimation of physical variables given relatively limited observational data. 
There has been a proliferation in machine learning-based approaches to predict physical quantities from observational data across various fields, including solar physics \citep{2019A&A...626A.102A}, interstellar medium \citep{2019ApJ...882L..12P,2020ApJ...905..172X,2020ApJ...890...64X,2022ApJ...926...19X,2022ApJ...941...81X}, and in the realm of galaxies and cosmology \citep{2019MNRAS.484.4683W,2022MNRAS.511.3446N}. 
% Mechanical feedback, encompassing protostellar outflows and stellar wind-driven bubbles, leaves discernible features on the morphology of the molecular clouds. 
Machine learning provides a powerful tool to study mechanical stellar feedback as it enables complex morphological features, previously only detectable by visual inspection, to be identified quickly and reliably. Recent studies have developed and employed a deep learning method called \CASI\ (Convolutional Approach to Structure Identification) to systematically identify protostellar outflows and wind-driven bubbles in nearby molecular clouds using molecular line data cubes \citep{2019ApJ...880...83V,2020ApJ...905..172X,2020ApJ...890...64X,2022ApJ...926...19X}. 
%Machine learning provides a promising avenue for improving the estimation of physical variables given relatively limited observational data. %This can be particularly valuable in tackling challenging tasks such as predicting radiation feedback.

Recently Denoising Diffusion Probabilistic Models (DDPMs) have emerged as powerful and reliable tools for image generation \citep{pmlr-v37-sohl-dickstein15,NEURIPS2020_diffusion}, which have shown great potential in addressing prediction tasks within the field of astronomy. \citet{2022MNRAS.511.1808S} employed DDPMs to generate synthetic images resembling observed galaxies, achieving a high level of realism. In another study, \citet{2023arXiv230509121W} utilized DDPMs to enhance image quality and suppress noise in interferometric observations. Furthermore, \citet{2023ApJ...950..146X} applied DDPMs to infer the number density of molecular clouds—a parameter notoriously challenging to measure based on readily obtainable column density maps. DDPMs exhibit superior accuracy in predicting molecular cloud number density, underscoring their effectiveness and reliability in the estimation task.

In this paper, we employ a deep learning approach based on DDPMs to estimate the radiation field strength induced by massive stars within molecular clouds. Specifically, we utilize multiple bands of dust emission to infer the radiation field strength. In Section~\ref{Data and Method}, we elucidate the diffusion model utilized in our analysis and delineate the procedure employed to generate the training set from magnetohydrodynamic (MHD) simulations. Subsequently, in Section~\ref{Results}, we comprehensively evaluate the performance of our diffusion model in predicting the radiation field strength. Additionally, we apply our diffusion model to actual observational data, as detailed in Section~\ref{Results}. Finally, we consolidate our findings and draw conclusions in Section~\ref{Conclusions}.

\section{Data and Method}
\label{Data and Method}

\subsection{Magnetohydrodynamics Simulations}
\label{Magnetohydrodynamics Simulations}

We employ MHD simulations acquired from the STARFORGE project \citep[STAR FORmation in Gaseous Environments,][]{2021MNRAS.506.2199G}. The project introduces a novel numerical framework for conducting three-dimensional radiation MHD simulations of star formation, allowing for a comprehensive examination of multiple processes. These processes encompass the formation, accretion, evolution, and dynamics of individual stars within massive giant molecular clouds (GMCs), while considering the intricate effects of stellar feedback. Stellar feedback mechanisms taken into account include jets, radiative heating and momentum, stellar winds, and supernovae.

The simulations in the STARFORGE project utilize the GIZMO code \citep{2015MNRAS.450...53H}, which incorporates the mesh-free Lagrangian MHD (MFM) method. Specifically, the star cluster formation is simulated within a GMC characterized by an initial mass of 2$\times 10^{4}$ \msun\ and a radius of 10 pc \citep{2022MNRAS.512..216G,2022MNRAS.515.4929G}. The initial magnetic field strength is set to 2 $\mu$G, and the cloud possesses an initial virial parameter of 2. The simulations achieve a mass resolution of $10^{-3}$ \msun\ and span an evolutionary timeframe of approximately 9 Myr. The interstellar radiation field (ISRF) default configuration is scaled to the background SED of the Solar neighborhood, with the \citet{draine_1978_isrf} value of $G_{0}$ = 1.7 in the FUV band. Additionally, we employ simulations where the ISRF is intensified by a factor of 10 and 100, corresponding to $G_{0}$ = 17 and $G_{0}$ = 170 \citep{2022MNRAS.515.4929G}. This alternative setup enables us to assess the performance of the machine learning model under stronger radiation field conditions. Significantly, the simulations account for stellar feedback by incorporating accretion- and fusion-powered stellar radiation in five distinct frequency bins. These bins include H-ionizing ($\lambda <912$ \r{A}), far ultraviolet (912 \r{A}$<\lambda <1550$ \r{A}), near ultraviolet (1550 \r{A}$<\lambda <$3600 \r{A}), optical-to-near infrared (3600 \r{A}$<\lambda <3 \mu m$), and far infrared ($\lambda >3 \mu m$) ranges. It is important to highlight that our work incorporates simulations spanning different evolutionary stages, ranging from 2 Myr to 8 Myr. This wide temporal range encompasses both early and late stages of star formation as well as the evolution of GMCs.

In addition to the full physics simulations, we incorporate a specific simulation that emphasizes the impact of stellar winds and radiation feedback while deactivating the presence of jets \citep{2022MNRAS.515.4929G}. This alternative simulation configuration introduces slight variations in the physical setup and enhances the diversity of cloud morphologies within the simulations. By including this simulation in our analysis, we expand the range of training data and further enrich the training set for our machine learning model. We provide a summary of the adopted simulations in Table~\ref{tab-simulation-STARGORGE}. For further comprehensive information regarding the STARFORGE project, additional details can be found in \citet{2021MNRAS.506.2199G}.

\subsection{MonR2 Observations}
\label{Monoceros R2 Observations}

%%%We utilize data obtained from the Spitzer and Herschel data archives to study the Monoceros R2 regions. The Spitzer observation from Program ID 20403 (Gutermuth et al., in preparation) provides the 4.5 \um\ observation captured by the Infrared Array Camera, while the 24 \um\ data is obtained from the Multiband Imaging Photometer. Additionally, we employ the 250 \um\ data from the SPIRE instrument on the Herschel satellite, as documented in \citet{2016MNRAS.461...22P}. The observational coverage encompasses an area of 4.$^{\circ}$30 by 4.$^{\circ}$36, centered at 06$^{h}$08$^{m}$46.$^{s}$90 RA (J2000), -06$^{\circ}$23$^{\prime}$12.$^{\prime\prime}$33 Dec (J2000). The spatial resolutions of the observations are 1.$^{\prime\prime}$8 for 4.5 \um, 6$^{\prime\prime}$ for 24 \um, and 25$^{\prime\prime}$ for 250 \um. The Monoceros R2 regions are located at a distance of 893$\pm$40 pc \citep{2016ApJ...826..201D}. To maintain physical scale consistency with our training set, we convolve all three-band images to match a resolution of 29$^{\prime\prime}$.

%%%\DX {Rob, would you mind providing additional details on this? Thank you. }
%%{Rob's write-up; I think you can delete the stuff above that I didn't realize existed when I wrote this:} 

%To test the trained model on real data, 
We adopt the Monoceros R2 (MonR2) GMC as an observational test case. MonR2 is well observed at all bandpasses of interest for this project. MonR2 is located 860~pc away, is 33,000 M$_{\odot}$, and hosts over 900 young stellar objects with excess IR emission indicative of dusty circumstellar material such as protoplanetary disks or protostellar envelopes \citep{pokhrel2020}. Thus, our fiducial STARFORGE calculation provides a reasonable representation of the MonR2 region given its cloud mass, evolutionary stage and level of star formation activity.

We adopt SESNA \citep[{\it Spitzer} Extended Solar Neighborhood Archive; Gutermuth et al. in prep.;][]{pokhrel2020} {\it Spitzer} \citep{werner2004} mosaics at 4.5~$\mu$m from the Infrared Array Camera \citep[IRAC;][]{fazio2004} instrument and 24~$\mu$m from the Mid-Infrared Photometer for {\it Spitzer} \citep[MIPS;][]{rieke2004} instrument. For the 250~$\mu$m image, we use the {\it Herschel} \citep{pilbratt2010} Spectral and Photometric Imaging REceiver \citep[SPIRE;][]{griffin2010} image from \citet{pokhrel2016} that included an absolute calibration correction to Planck High Frequency Instrument \citep[HFI;][]{planck2011} data of the same region of sky.  

The trained model as described in \S\ref{Synthetic Dust Observations} operates on physical scales of 1/8 parsec per pixel, which translates to 30$^{\prime\prime}$ per pixel at MonR2's distance. For our analysis we resample all three infrared images to a common pixel grid set by the IRAC 4.5~$\mu$m image, the highest resolution data of the collection at 2.$^{\prime\prime}$2 beam width and 0.$^{\prime\prime}$87 pixel size (MIPS 24~$\mu$m is 6.$^{\prime\prime}$3 beam width and 1.$^{\prime\prime}$8 per pixel; SPIRE 250~$\mu$m is 18$^{\prime\prime}$ beam width and 6$^{\prime\prime}$ per pixel). Since the beam resolutions of all three images are less than our final pixel scale, we simply box-average and down-sample the flux into the desired 30$^{\prime\prime}$ pixel size grid. We then apply a mask to limit consideration to those pixels with coverage in all three bandpasses. The resulting coverage spans an area of 5.23 deg$^2$. This treatment was applied using standard routines for these tasks (e.g., {\it hastrom, hrebin}) from the IDL Astronomy User's Library \citep{landsman1993}.      

\subsection{Synthetic Dust Observations}
\label{Synthetic Dust Observations}

\begin{figure*}[hbt!]
\centering
\includegraphics[width=0.98\linewidth]{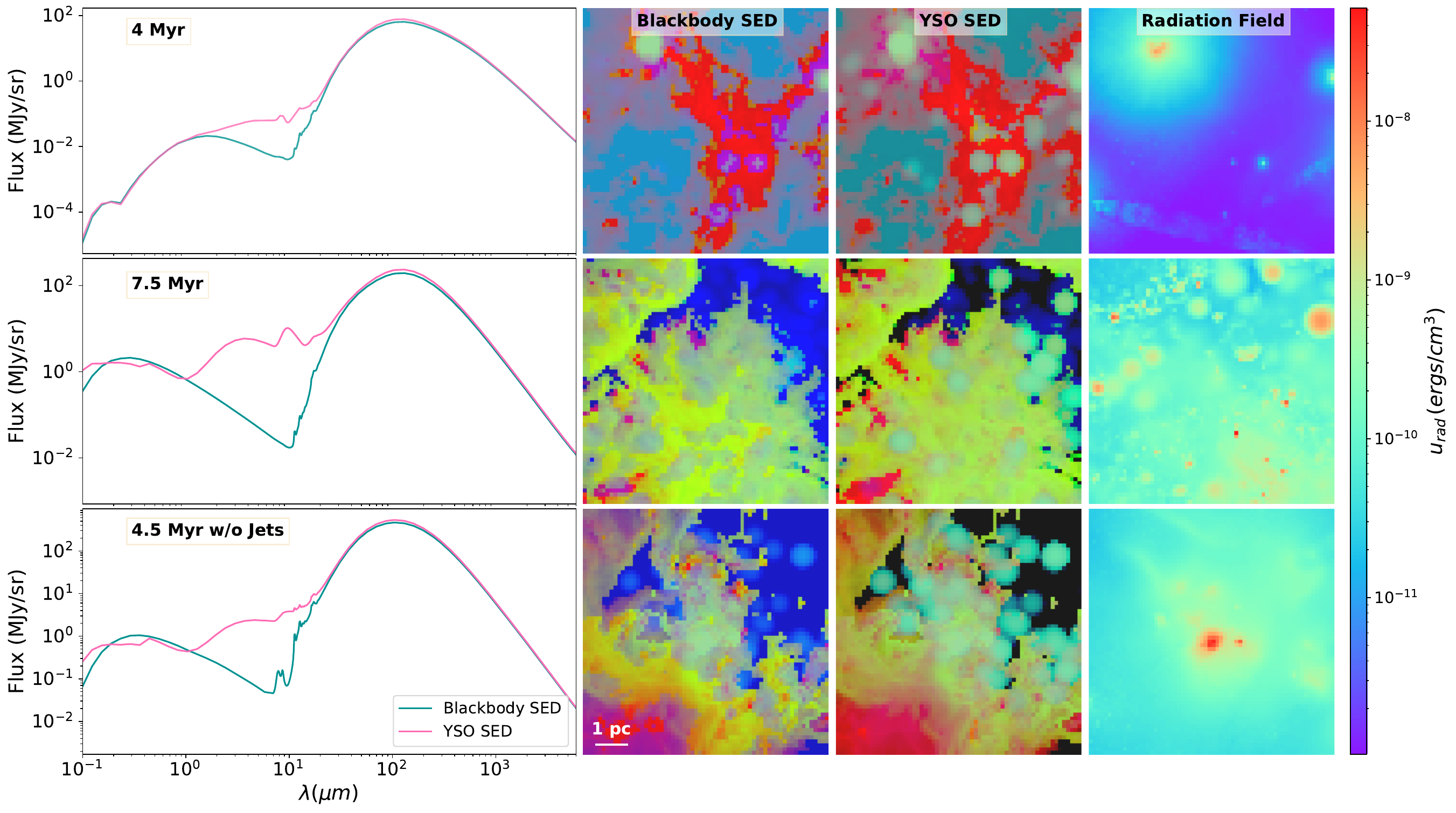}
\caption{Synthetic dust observations, including images and SEDs, for simulations at various evolutionary stages and with different feedback configurations. The first column illustrates the SEDs of the synthetic observations. The second and third columns showcase the three-color synthetic dust images at 4.5 \um\ (blue), 24 \um\ (green), and 250 \um\ (red) wavelengths, utilizing different radiative transfer configurations. %{mike: what are the round blue features in the YSO SED, and why do they truncate sharply to black?}. 
The fourth column presents the projected radiation field %{mike: what's the weighting on this projection? by volume?} 
strength obtained from the simulations, measured in ergs/\cmc.}
\label{fig.synthetic_dust_sed_example_1}
\end{figure*}

To calculate the dust temperature and generate synthetic dust emission at multiple wavelengths, we employ the 3D radiative transfer code \radmc\ \citep{2012ascl.soft02015D}. STARFORGE uses a subgrid model for protostellar evolution \citep{2009ApJ...703..131O} and stores the luminosity, radius, and effective temperature of each source. Due to computational constraints, it is not feasible to assign a unique stellar spectrum to each individual star. Instead, we categorize the stars into four groups based on their effective temperature: $<$2000 K, 2000-5000 K, 5000-10000 K, and $>$10000 K. For each category, we calculate the mean effective temperature by considering the mean luminosity and mean surface area of the stars within that category. The calculation of the mean effective temperature is solely based on the stars that are located within the domain where the radiative transfer is conducted. An example of star categorization in one simulation snapshot is presented in Appendix~\ref{Star Categorization in Radiative Transfer}.

We explore two different approaches for modeling the stellar spectrum, specifically the spectral energy distribution (SED), in our study: the ``Blackbody SED" and the ``YSO SED." The Blackbody SED assumes a blackbody spectrum based on the mean effective temperature of the stars. However, circumstellar disks play a significant role in shaping the SEDs of young sources \citep{2003ApJ...598.1079W,2007ApJS..169..328R,2012ApJ...753...98O}. In the STARFORGE simulations, the formation of circumstellar disks is suppressed due to strong magnetic braking.  The YSO SED accounts for the emission reprocessing (e.g. extinction, absorption and remission, scattering) caused by these (missing) disks. We adopt the stellar spectra with disks from \citet{2017A&A...600A..11R}. For each category of stars, we retrieve the SED from the table in \citet{2017A&A...600A..11R} by selecting the one that closely matches the effective temperature and stellar radius of the star.% {mike: is there some age evolution in this model? if not, is it valid to assume the hot stars always have disk/envelope? if not, they will radiate at short wavelengths and there will be much more hot dust} 
The table also includes different inclination angles for young stars with disks. In our approach, we adopt the spectrum with an inclination angle that is closest to 45\deg.

\begin{figure*}[hbt!]
\centering
\includegraphics[width=0.78\linewidth]{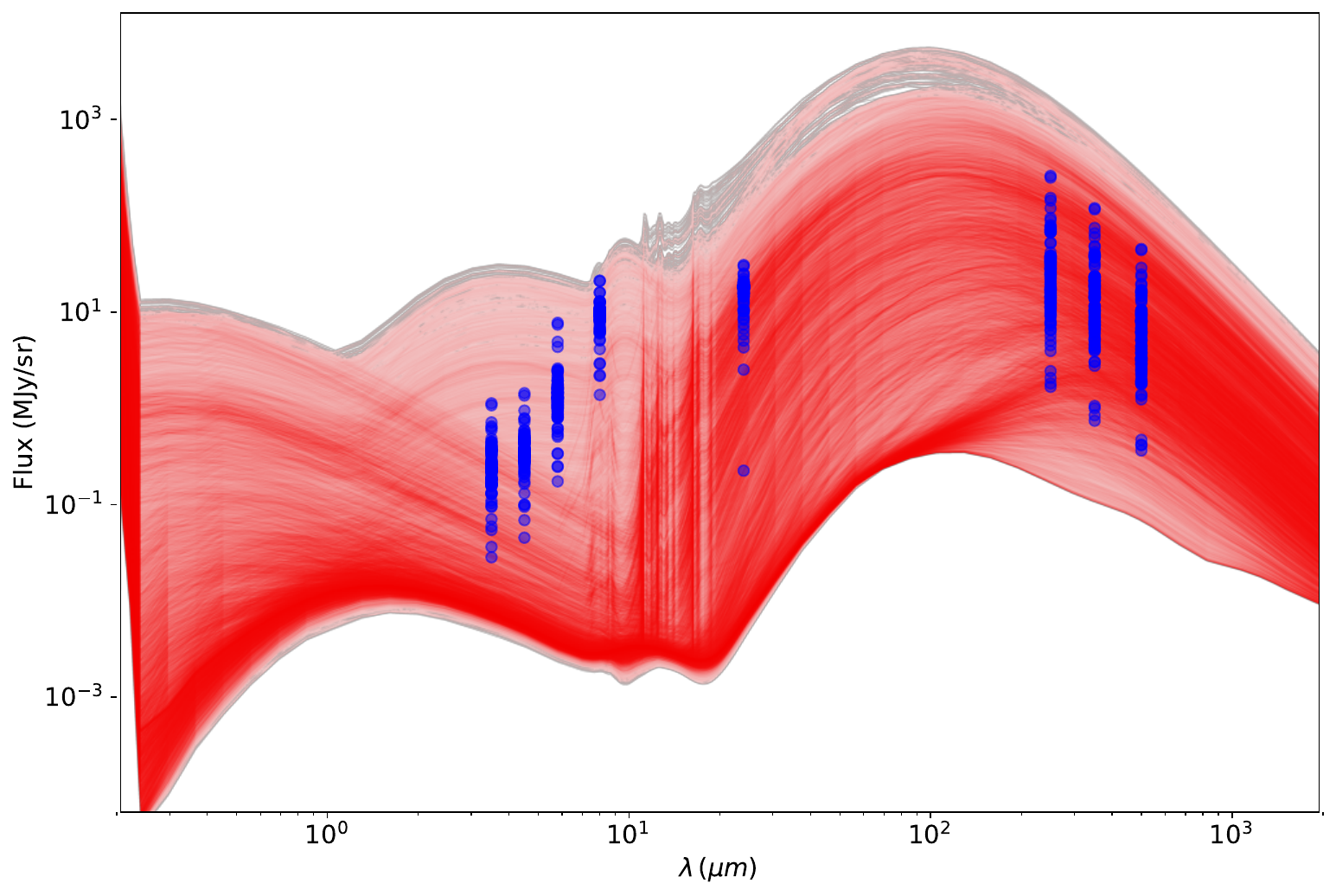}
\caption{A collection of all the synthetic dust SEDs, where the intensity of color is proportional to the number of stacked SEDs. Blue dots represent observations in {\it Spitzer} bands (3.5 \um, 4.5 \um, 5.8 \um, 8.0 \um, 24 \um) and {\it Herschel} bands (250 \um, 350 \um, 500 \um) within MonR2.%{mike: is this every snapshot of the simulation? Different viewing angles? Does this control for normalization in some way?}%% {Is this all of them? For all sources? How have you curated this 'collection'?}
}
\label{fig.synthetic_sed_all_MonR2}
\end{figure*}

\begin{table*}
\begin{center}
\caption{Summary of STARFORGE Simulations$^a$ \label{tab-simulation-STARGORGE}}
\begin{tabular}{ccccc}
\hline
\hline
      & $t_{s}$ (Myr) & ISRF($G_{0}$) & Jets & $N_{\rm sample}$\\
\hline
Training \&      & 3.5-7.5 & 1.7 & Yes & 6750\\
    Testing & 4-5.5 & 1.7 & No  & 3000\\
\hline
 \multirow{3}{*}{Testing}    & 5   & 17    & Yes & 81 \\
  & 5  & 170   & Yes & 81 \\
  & 1.7-4.3$^{b}$  & 1.7   & Yes & 127 \\
  \hline
\multicolumn{5}{p{0.5\linewidth}}{Notes:}\\
\multicolumn{5}{p{0.5\linewidth}}{$^a$ Evolutionary time, ISRF, whether protostellar jets are included, and the number of image samples.

$^b$ These simulations are initially subjected to turbulent driving for two crossing times, equivalent to 17.5 Myr. Furthermore, they incorporate the updated heating and cooling treatments. }

\end{tabular}%
\end{center}
\end{table*}%

% \begin{table*}
% \begin{center}
% \caption{Summary of STARFORGE Simulations$^a$ \label{tab-simulation-STARGORGE}}
% \begin{tabular}{ccccccccc}
% \hline
% \hline
%     &Label & $M(M_{\odot})$ & $R$(pc) & $\alpha$ & $B(\mu {\rm G})$  & $t_{s}$ (Myr) & ISRF($G_{0}$) & Jets \\
% \hline
% Training\&    & 1     & 2$\times 10^{4}$ & 10    & 2     & 2     & 3.5-7.5 & 1.7 & Included \\
%     Testing & 2     & 2$\times 10^{4}$  &  10     &    2     &
%     2   & 4-5.5 & 1.7 & No \\
% \hline
%  \multirow{2}{*}{Testing}    & 3     &  2$\times 10^{4}$     &    10   &   2      & 2     & 5   & 17    & Included \\
%   & 4     &    2$\times 10^{4}$   &  10     &      2   & 2 & 5  & 170   & Included \\
%   \hline
% \multicolumn{9}{p{0.97\linewidth}}{Notes:}\\
% \multicolumn{9}{p{0.97\linewidth}}{$^a$ Model label, initial cloud mass, initial cloud radius, virial parameters, initial magnetic field strength, evolutionary stages, ISRF and whether including protostellar jets.}

% \end{tabular}%
% \end{center}
% \end{table*}%

In the radiative transfer calculation, we employ two different dust models depending on the gas number density. For gas number densities exceeding $10^5$ \cmc, we utilize the dust model proposed by \citet{2017ApJS..233....1K} for dense gas. This model consists of three dust compositions: 80.63\% big grains ($>$200 \AA), 13.51\% very small grains (vsg, 20-200 \AA), and 5.86\% ultrasmall grains (usg, $<$20 \AA) in the form of polycyclic aromatic hydrocarbon (PAH) molecules. On the other hand, for gas number densities below $10^5$ \cmc, we adopt the dust model developed by \citet{2023ApJ...948...55H} specifically designed for diffuse gas. This model incorporates two dust components: astrodust (90.69\%) and PAH (9.31\%). In Appendix~\ref{Exploration of Different Dust Models}, we investigate various cutoffs on gas number densities when selecting dust models, as well as explore different dust models. This exercise illustrates that the choice of dust model is crucial to reproduce the observed SEDs; our hybrid model reproduces the relative fluxes in the three bands significantly better than the canonical \citet{1984ApJ...285...89D} model or either of the two models alone.

Dust heating in molecular clouds is influenced by multiple mechanisms, with radiation from stars and the ISRF playing dominant roles. %%However, viscosity and turbulent/shock dissipation can also contribute significantly, particularly in high-velocity jets. The STARFORGE simulation considers these mechanisms and calculates gas and dust temperatures separately, but adopts a relatively simplified treatment for dust \citep{2021MNRAS.506.2199G}. 
It is important to mention that the simulation data used in this study does not include the saved dust temperature during the simulation runs. %Only the gas temperature is available for analysis. 
The gas temperature is not a good proxy for the dust temperature, as they differ by an order of magnitude in shocks and in lower density regions, where the dust are gas are not well-coupled. Consequently, using the gas temperature in place of the dust temperature in the radiative transfer 
%would lead to a significant discrepancy, with an order of magnitude difference between the two temperatures. This discrepancy 
would result in a substantial difference in the calculated dust emission, spanning several orders of magnitude. To address this, we utilize the \radmc\ package to calculate the dust temperature in post-processing.
Given that GIZMO utilizes a Lagrangian meshless finite mass method rather than a Cartesian grid, we employ the $yt$ toolkit \citep{2011ApJS..192....9T} to sample the simulation data and transform it into a uniform Cartesian grid. We use this processed data as the input for \radmc\ in our analysis. We assume a gas-to-dust ratio of 100 and incorporate the Henyey-Greenstein anisotropic scattering model in the radiative transfer calculation. 

Due to computational constraints, we generate the synthetic dust images for each $10\times10\times10$ pc$^3$ box with an image resolution of 80$\times$80 pixels. We verify that the radiative transfer results remain robust regardless of the resolution. In Appendix~\ref{Exploration of Different Dust Models}, we present the radiative transfer simulations with a resolution of 256$\times$256, and we show  that the resulting SEDs are consistent with those obtained at lower resolutions. %% {mike: Do we understand roughly how the RT results depend on the size of the box? e.g. 20x20x20 to enclose the whole cloud?}. 

To account for stars located near the box boundaries, we apply an additional post-processing step. After generating the initial synthetic dust images with dimensions of 80$\times$80 pixels, we crop a 1 pc boundary on all four sides of the image. This results in a final image size of 64$\times$64 pixels, representing an $8\times8$ pc$^2$ sky area.

Figure~\ref{fig.synthetic_dust_sed_example_1} illustrates the synthetic dust emission at 4.5 \um, 24 \um, and 250 \um, considering the two different treatments for the stellar spectrum, a blackbody SED and a YSO SED, at various evolutionary stages. The figure also presents the projected radiation field averaged by the radiation energy along the line-of-sight. The synthetic dust emission SEDs are sensitive to the choice of stellar spectrum. %specifically between blackbody SEDs and YSO SEDs. 
The SEDs generated with blackbody SEDs as the radiating sources exhibit two distinct peaks, representing the contributions of stellar radiation in the optical to near-infrared range and dust emission of the cloud material in the mid- to far-infrared range. In contrast, the SEDs generated with the radiating sources modeled as YSOs display an infrared excess from 1-10\,$\mu$m. This difference arises 
%from the disparity between YSO SEDs and blackbody SEDs, where 
because the YSO SEDs exhibit higher levels of infrared emission due to emission reprocessing by the dust in circumstellar disks. 
%SO Alternative text:
We note that neither set of synthetic images exhibit outflow features, which often appear in these bands \citep{2007ApJ...670L.131L,2008ApJ...679.1364T,2010ApJ...720..155T}. Some of the excess observed emission likely arises from shock-excited H$_2$ and CO lines \citep{2008AJ....136.2391C}, which are not included in our radiative transfer modeling step.
%%%In contrast, the synthetic dust SEDs generated using blackbody SEDs and those incorporating the internal heat source exhibit remarkable similarity. This is achieved by reducing the gas thermal energy by a factor of $10^{4}$ as internal heat sources during the radiative transfer, ensuring minimal changes to the SEDs while emphasizing the emission from high-velocity shocked gas. The synthetic dust maps with internal heat sources illuminate the jet cavities, producing regions characterized by relatively high temperatures. In cases where no prominent jets are observed, such as in the synthetic dust emission from simulations without jets, the dust emission maps exhibit similarities between those generated with blackbody SEDs and those with blackbody SED plus the internal heat source.

\subsection{Constructing the Training Set}
\label{Constructing the Training Set}

In this study, we utilize three specific bands of dust emission, namely 4.5 \um, 24 \um, and 250 \um, as input for training the machine learning model to predict the radiation field at the pixel level. These bands cover both near-infrared and far-infrared dust emission and, importantly, encompass information that is well-modeled by our training set. Figure~\ref{fig.synthetic_sed_all_MonR2} presents a gallery of SEDs for all the synthetic data, with darker colors indicating a higher number of stacked SEDs. Observations from the MonR2 GMC are included in the figure for reference, showcasing the {\it Spitzer} bands (3.6 \um, 4.5 \um, 5.8 \um, 8.0 \um, 24 \um) and {\it Herschel} bands (250 \um, 350 \um, 500 \um). 
%these observations are described in \sectionmark \ref{Monoceros R2 Observations} below). 
The synthetic SEDs demonstrate a broad range of coverage for the observed data points. Nonetheless, there is an evident discrepancy in the synthetic SEDs, particularly in the 8 \um\ emission, where it is noticeably underestimated. This discrepancy is caused by %to the presence of 
strong PAH emission, indicating that the adopted dust model does not replicate the observed PAH emission adequately. Part of the discrepancy is likely due to the absence of non-thermal excitation mechanisms, such as shocks, which are not included in post-processing. It is important to note that the presence of \h2\ lines at 2.22 \um, 2.41 \um, 2.63 \um, and 3.00 \um\ may cause contamination in the 3.6 \um\ bandpass. Additionally, the strong aromatic infrared bands at 3.30 \um, 6.20 \um, 7.70 \um, 8.60 \um, 11.30 \um, and 12.70 \um\ can potentially contaminate the 3.6 \um, 5.8 \um, and 8 \um\ bands \citep{2019A&A...632A..84F}. As our project aims to infer the ISRF using a limited amount of data, it is crucial to ensure that the synthetic data closely resembles the real data. To mitigate this issue and achieve better performance, we have excluded bandpasses that include \h2\ lines and strong PAH feature emission (3.6 \um, 5.8 \um, 8.0 \um) in this study.

To enhance the model's capability to handle real observational data, which may include the presence of foreground and background stars, we randomly introduce bright %blobs and dots 
false sources that simulate such stars in the 4.5 \um ~images. These sources exhibit a two-dimensional Gaussian intensity distribution, where the peak intensity is randomly selected as a fraction between 0.1 and 1 of the 99.5th percentile of the 4.5 \um ~images. For each synthetic dust map, we utilize the projected radiation field averaged along the line-of-sight, weighted by radiation energy, as the target for the machine learning training. In summary, the input consists of three-band dust emission maps ($3\times 64\times 64$), while the target or model output is the corresponding projected radiation field ($64\times 64$).

We generate a total of 9,750 synthetic dust maps, encompassing various evolutionary stages, feedback configurations, and treatments for the stellar spectrum. To evaluate the performance of the machine learning model, the data is randomly split into an 80\% training set and a 20\% test set.

To comprehensively assess the performance of the machine learning model across different environments, we generate an additional 162 synthetic dust observations. These synthetic observations are generated from simulations with an ISRF strength of $G_{0}$ = 17 and $G_{0}$ = 170. Importantly, these new synthetic dust observations are entirely distinct from the training data, allowing us to evaluate the model's performance on previously unseen data.

\subsection{Denoising Diffusion Probabilistic Models}
\label{Denoising Diffusion Probabilistic Models}

\begin{figure*}[hbt!]
\centering
\includegraphics[width=0.99\linewidth]{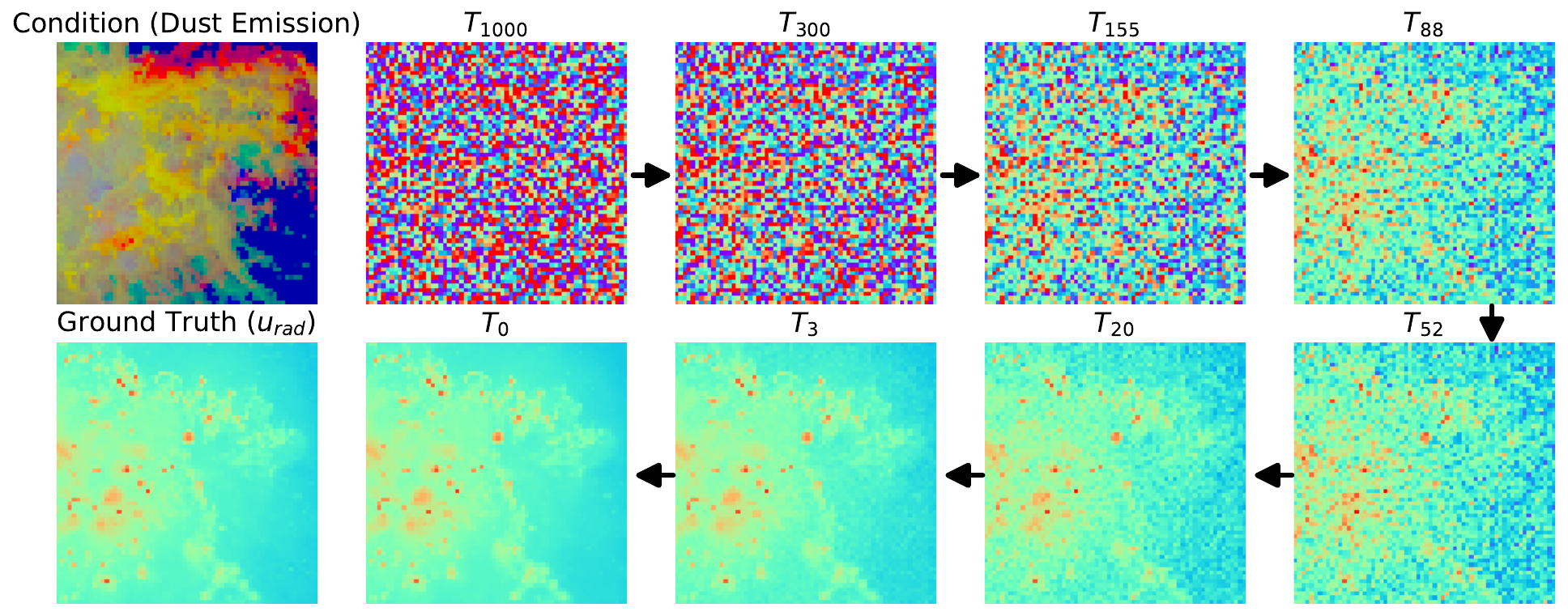}
\caption{Demonstration of the diffusion process (reverse) on a sample in the test set. In the upper row, the first panel represents the input (condition) for the diffusion model, while in the lower row, the first panel represents the corresponding target (ground truth). The initial status is denoted as $T_{1000}$, which corresponds to the random Gaussian noise. The final states of the reverse Markov chain, representing the final predictions by the diffusion model, are indicated as $T_{0}$. The intermediate steps of the reverse Markov chain, ranging from $T$ = 0 to 1000, are depicted in the remaining panels. %%% {State what is in each of the panels -- what is $T_n$.} 
}
\label{fig.diffusion-demo-example}
\end{figure*}

Diffusion models, also known as denoising diffusion probabilistic models (DDPMs), are state-of-the-art generative methods used in deep learning and computer vision research \citep{pmlr-v37-sohl-dickstein15,NEURIPS2020_diffusion,stablediff}. These models leverage probability theory and stochastic processes to effectively model and reconstruct data, with a focus on modeling the conditional distribution of clean data given noisy observations. By estimating the underlying distribution of the data, DDPMs capture its statistical properties, patterns, variations, and complexities.

The primary objective of DDPMs is to denoise and reconstruct the original signal from noisy or corrupted data. By modeling the distribution of clean data and incorporating diffusion processes, DDPMs excel at recovering true underlying structure while suppressing noise. The diffusion process is a key component, governing the evolution of the data distribution over time. 

The DDPM starts with a simple initial distribution and gradually transforms it into the target distribution, which represents the conditional distribution of clean data. This transformation occurs through a sequence of diffusion steps, involving diffusion and denoising operations. Controlled noise is introduced during diffusion to guide the data along a diffusion path, followed by a denoising step that estimates the clean data from the noisy observations. Typically, deep neural network architectures like CNNs are employed in the denoising step, training them to map noisy observations to clean data. Figure~\ref{fig.diffusion-demo-example} illustrates an example of the reverse process applied to our test data, demonstrating the gradual conversion of Gaussian noise into our desired target over 1000 time steps.

In our work, we adopt the same diffusion model described in \citet{2023ApJ...950..146X}, where a detailed mathematical explanation of DDPM formulation is provided. To train our diffusion model for the task of reconstructing the radiation fields based on three-band dust emission maps, we follow the training strategy outlined in \citet{2023ApJ...950..146X}.

\section{Results}
\label{Results}

\subsection{Assessing the Performance of the Diffusion Model}
\label{Assessing the Performance of the Diffusion Model}

\begin{figure*}[hbt!]
\centering
\includegraphics[width=0.99\linewidth]{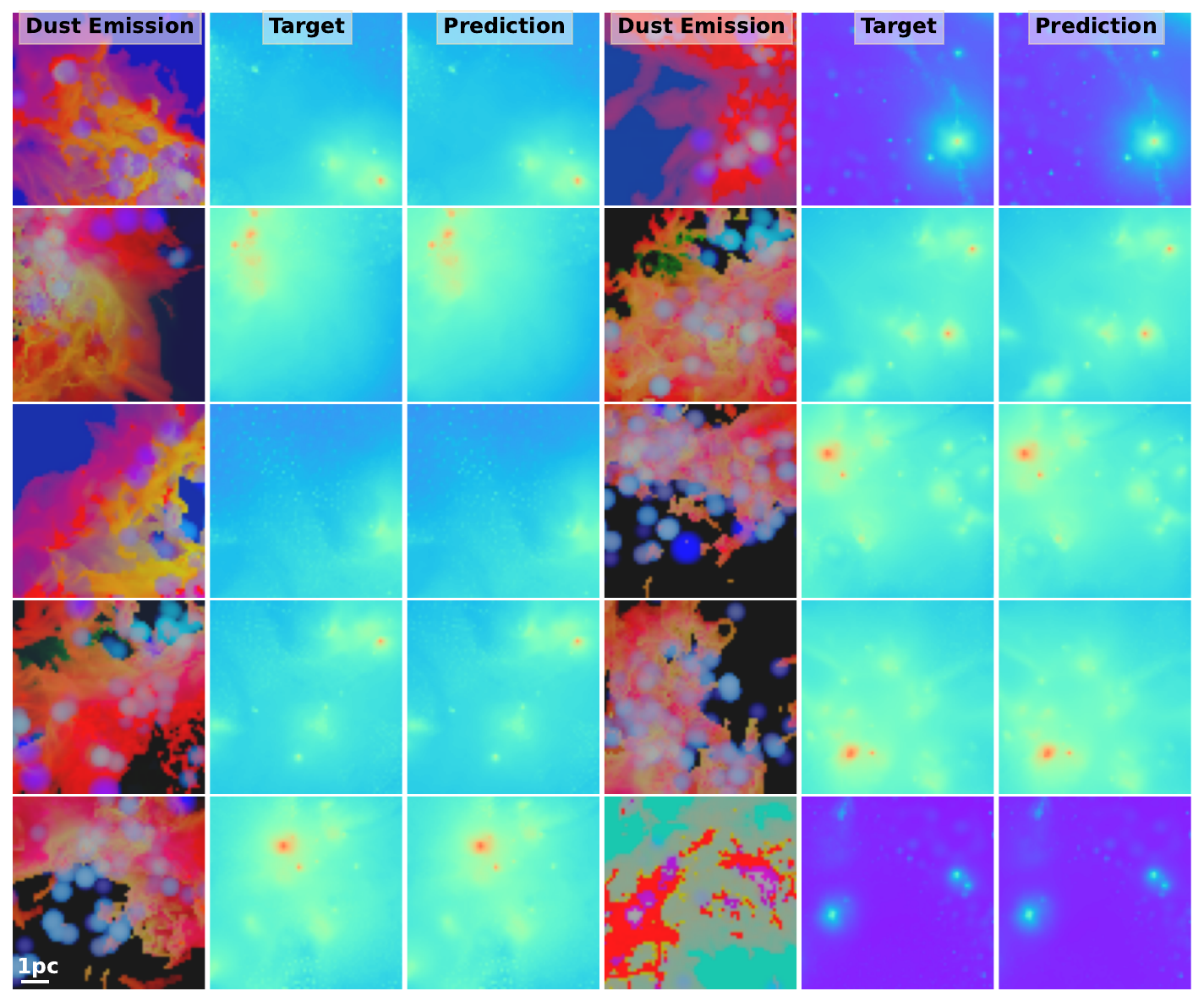}
\caption{Predicted radiation field generated by the diffusion model (columns 3 and 6), along with the corresponding three-band dust emission at 4.5 \um\ (blue), 24 \um\ (green), and 250 \um\ (red) (columns 1 and 4), and the ground truth radiation field (columns 2 and 5).}
\label{fig.diffusion_test_dust_img_example_1}
\end{figure*} 

\begin{figure*}[hbt!]
\centering
\includegraphics[width=0.69\linewidth]{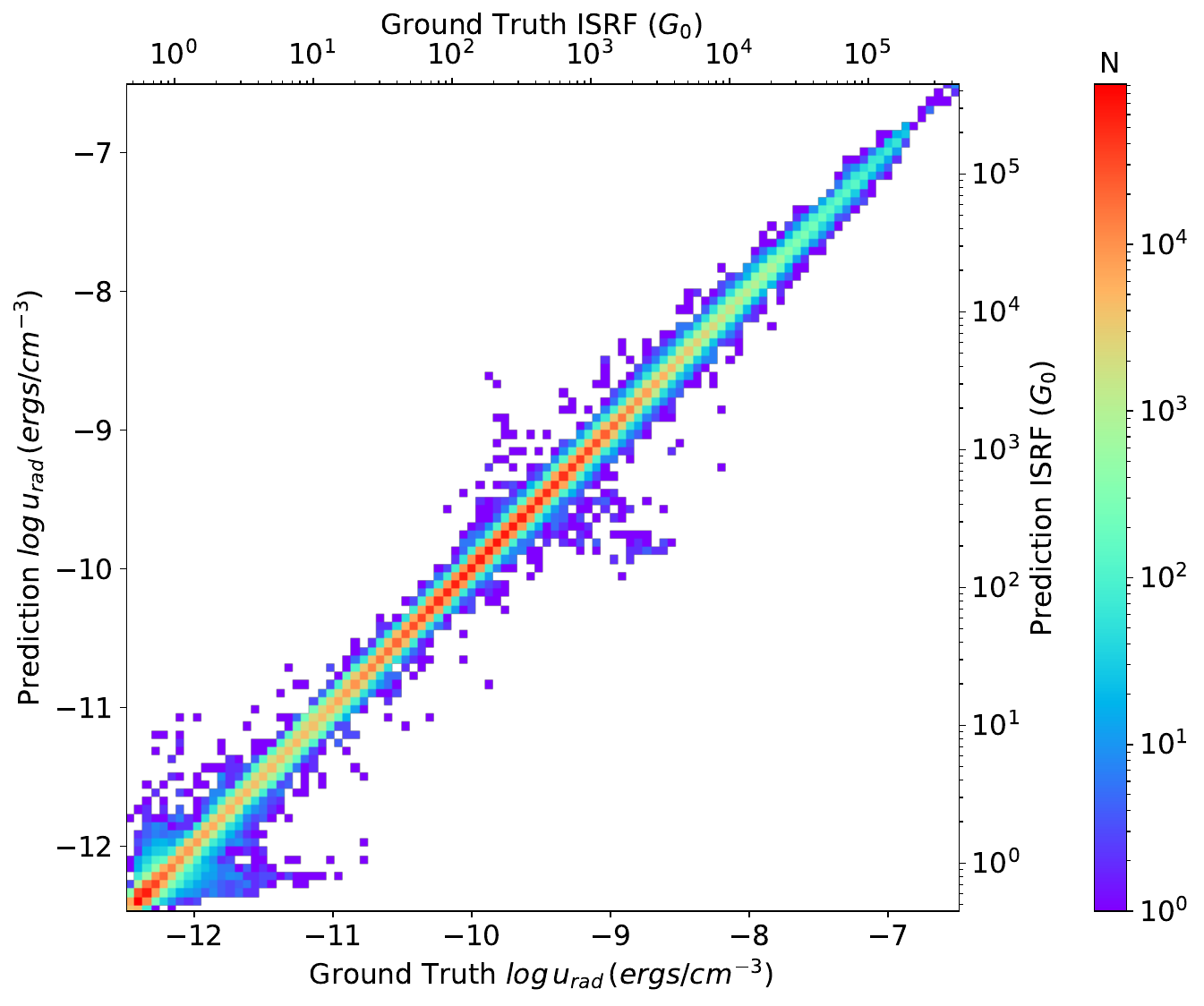}
\caption{2D histogram illustrating the correlation between the predictions of the diffusion model and the ground truth values of the radiation fields.}% The bernard model should work OK in the outer parts of the cloud  where $ISRF \sim 1$...}
\label{fig.diffusion_test_hist_1}
\end{figure*}

\begin{figure}[hbt!]
\centering
\includegraphics[width=0.98\linewidth]{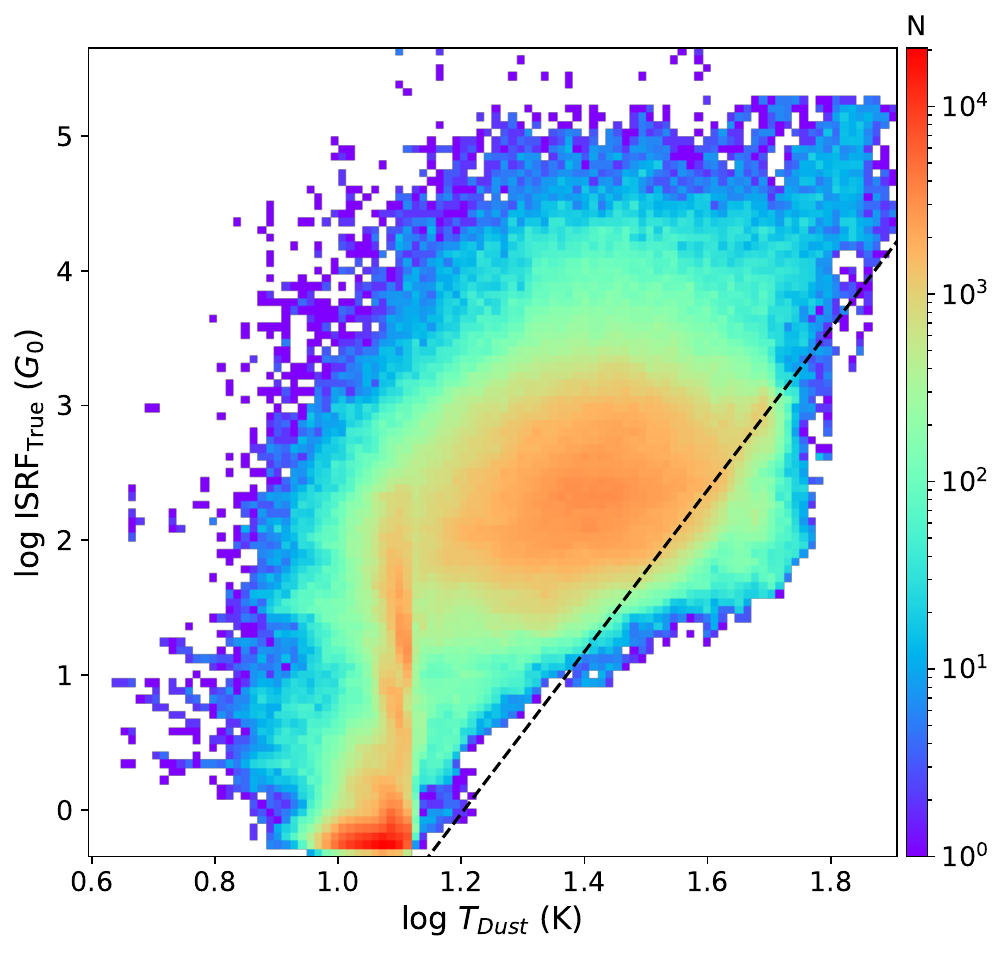}
\caption{2D histogram illustrating the correlation between the true ISRF and the dust temperature in the synthetic test data. The black dashed line represents the relationship predicted by Equation~\ref{equation-isrf-dust} \citep{2010A&A...518L..88B}, where the dust emissivity index $\beta$ is assumed to be 2 and the incident SED is that of the unattenuated local ISRF. %%{mike: this is off by a factor of about 100 on the y axis I think, due to the units issue.}
%points where the ISRF$_{\rm T-Dust}$ equal ISRF$_{\rm True}$. ISRF$_{\rm T-Dust}$ is derived using the formula $(\frac{T_{\rm Dust}}{17.5\, \rm K})^{4+\beta}$ from \citet{2010A&A...518L..88B}, where the dust emissivity index $\beta$ is set to 2 in our analysis.
}
\label{fig.diffusion_test_hist_dust_pred}
\end{figure}

In this section, we assess the performance of the diffusion model in predicting the radiation field strength based on the three-band dust emission. We begin by evaluating the diffusion model's performance on the test set. Figure~\ref{fig.diffusion_test_dust_img_example_1} illustrates the predicted radiation field, alongside the three-band dust emission and the ground truth radiation field. By visual comparison, the diffusion model accurately predicts the radiation field at the pixel level. 

To further evaluate the performance, we present a 2D histogram depicting the correlation between the diffusion model's predictions and the ground truth values of the radiation fields in Figure~\ref{fig.diffusion_test_hist_1}. The histogram demonstrates a strong alignment between the diffusion model's predictions and the ground truth radiation field values. The deviation between the true value and the predicted value is within a factor of 0.1. To provide a more interpretable representation of the radiation energy, we convert the radiation field energy into ISRF luminosity in the solar neighborhood. This conversion is achieved by adopting the mean intensity integrated over frequency from \citet{1983A&A...128..212M}, where $4\pi J = 0.0217 \,{\rm erg\, cm^{-2}\, s^{-1}}$. %%%We note that due to the extinction present in the molecular cloud, the projected radiation field intensity, averaged by radiation energy along the line-of-sight, may be significantly lower than the unity value observed in the solar neighborhood.

We further investigate the comparison between the traditional approach, which estimates the ISRF from dust temperature, and the actual ISRF in the test set. \citet{2010A&A...518L..88B} proposed an analytical formula to estimate the ISRF based on dust temperature in GMCs, following a power law:
\begin{equation}
\label{equation-isrf-dust}
{\rm ISRF}_{\rm T-Dust}=\left(\frac{T_{\rm Dust}}{17.5\, \rm K}\right)^{4+\beta},
\end{equation}
where the dust emissivity index $\beta$ is assumed to be 2, which is a good fit to our adopted dust model at long wavelengths. This equation follows from the balance of dust absorption and emission in the diffuse ISM, assuming an ambient ISRF with an SED like the ISRF in the Solar neighborhood. Therefore, it should describe the low-extinction parts of the cloud fairly well but not the inner parts subject to extinction and irradiation by protostars.

Figure~\ref{fig.diffusion_test_hist_dust_pred} depicts the correlation between the true ISRF and the dust temperature calculated using \radmc. We observe a weak or even unclear linear trend between these two quantities. Since the plot is in log scale, the pattern appears similar to the ISRF$_{\rm T-Dust}$ vs ISRF$_{\rm True}$ plot, but with different magnitudes in their values. %%{mike: trying to understand why the model } 
For reference, the one-to-one line of ISRF$_{\rm T-Dust}$ and ISRF$_{\rm True}$ is also shown in Figure~\ref{fig.diffusion_test_hist_dust_pred}. The inferred ISRF from the dust temperature exhibits a notable offset from the true ISRF, mainly due to the extinction of radiation from stars within the cloud by dust along the line-of-sight. This offset diminishes at low ISRF values, where line-of-sight extinction is minimal. Overall, this highlights a significant level of uncertainty, with scatter and offset of over a factor of 10, in the traditional approach for estimating the ISRF from dust emission as compared to our machine learning approach.
%In contrast, the machine learning approach demonstrates its effectiveness and capability in addressing these challenges.}
%SO I dont' think you need this last statement -- it is very general and doesn't add information.

{
\subsection{Testing on Out-of-Distribution Data}
\label{Testing on New Simulations}

Although the test set was not included in the training process, both the synthetic dust observations in the test set and the training set originate from the same sequence of MHD simulations. As a result, the diffusion model is potentially capable of learning all the intricacies within the synthetic dust emission and achieving unfairly accurate predictions, which could be significantly less accurate when applied to more diverse data. To address the possible issue of over-fitting, we evaluate the performance of the diffusion model on unseen data. This includes synthetic images created using different dust models and entirely novel MHD simulations featuring diverse physical parameters. {Since simulations can never perfectly model observations, even when great care is taken to match physical conditions and to include relevant physical effects, our training data is by definition out-of-distribution compared to the observational data. The tests presented here thus provide a more realistic assessment of the prediction accuracy of the model applied to observations.}

\subsubsection{Different Dust Models}

\begin{figure*}[hbt!]
\centering
\includegraphics[width=0.69\linewidth]{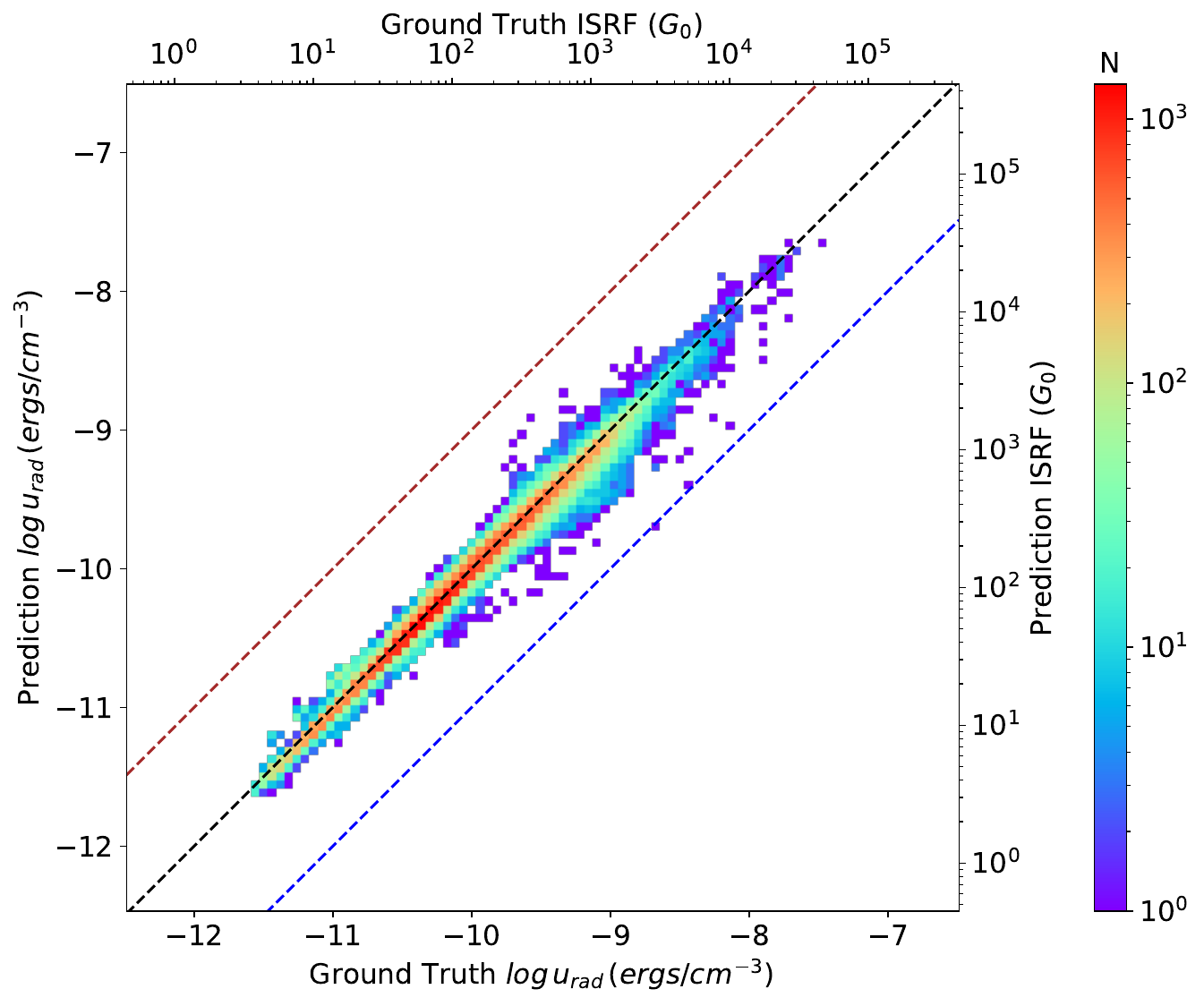}
\caption{Similar to Figure~\ref{fig.diffusion_test_hist_1}, but applied to synthetic dust images generated using different dust models. The black dotted line represents the one-to-one line, which represents perfect prediction. The blue dotted line represents the ten-to-one line, indicating underestimation by a factor of 10. The brown dotted line represents the one-to-ten line, indicating overestimation by a factor of 10. }
\label{fig.diffusion_test_hist_dustmodel}
\end{figure*} 

First, we evaluate the diffusion model's performance using synthetic dust images generated with alternative dust models, distinct from those used in the training set. It's important to note that the simulation data employed for this evaluation is identical to that used for generating the standard training and testing datasets. Therefore, the primary difference between the test data in this assessment and the standard training data lies in the selection of dust models. We consider two extreme scenarios, namely, the pure K17 model \citep{2017ApJS..233....1K} and the pure HD23 model \citep{2023ApJ...948...55H}.

Figure~\ref{fig.diffusion_test_hist_dustmodel} illustrates the correlation between the predictions of the diffusion model and the actual ground truth values of radiation fields in this evaluation. The histogram shows a strong alignment between the diffusion model's predictions and the actual radiation field values. The deviation between the true and predicted values is within a factor of 0.2, slightly larger than that observed in the fiducial test set described in Section ~\ref{Assessing the Performance of the Diffusion Model}. This suggests that the diffusion model can make robust predictions even when applied to the same sequence of MHD simulations with different dust model setups. Hence, the choice of a different dust model does not significantly impact the performance of the diffusion model during training and prediction.

\subsubsection{Higher ISRF}

\begin{figure*}[hbt!]
\centering
\includegraphics[width=0.99\linewidth]{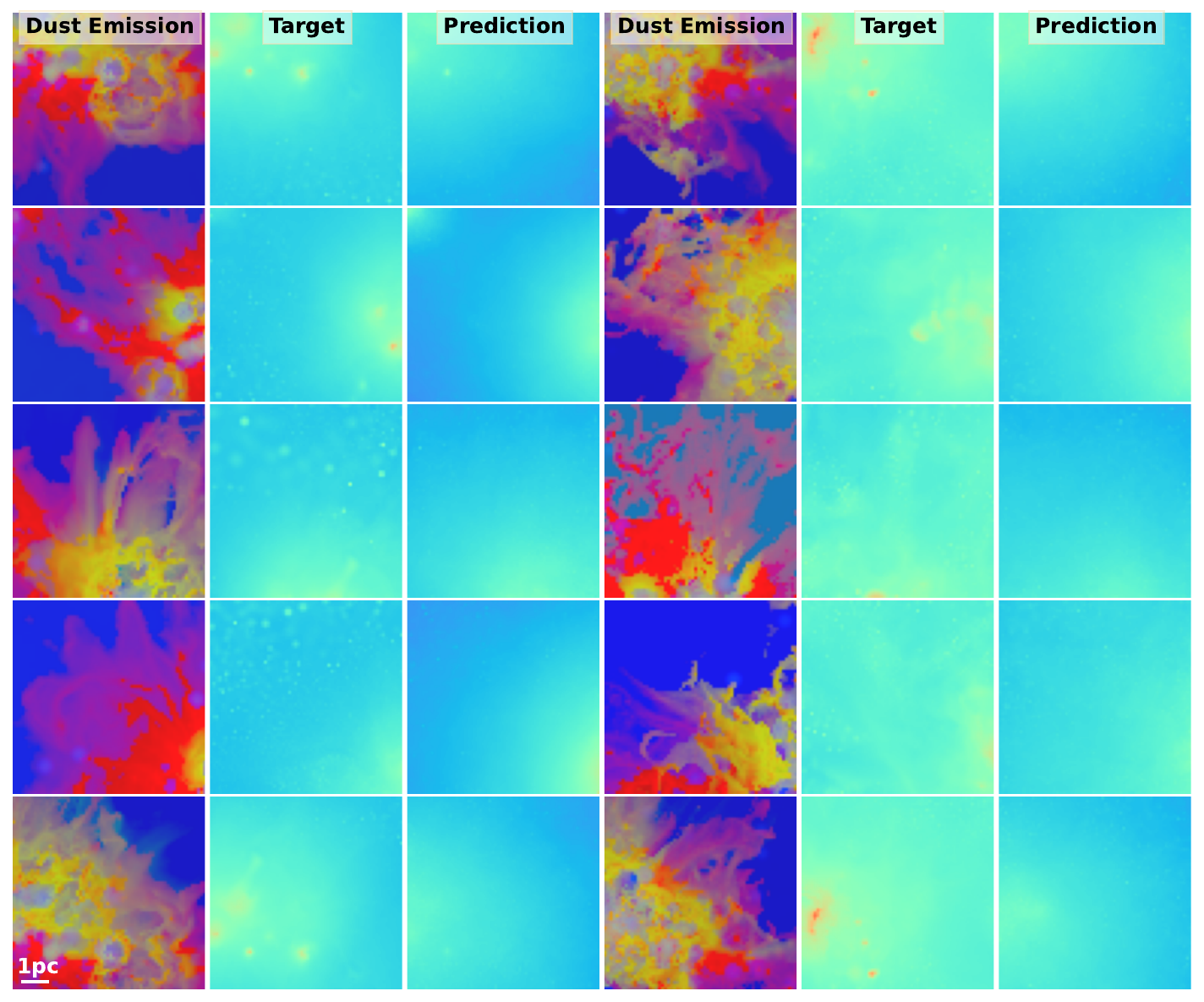}
\caption{Predicted radiation field generated by the diffusion model, along with the corresponding three-band dust emission and the ground truth radiation field. The left three panels correspond to new simulations with an ISRF of $G_{0}$ = 17, while the right three panels represent simulations with an ISRF of $G_{0}$ = 170.}
\label{fig.diffusion_test_dust_img_isrf10_1}
\end{figure*}

\begin{figure*}[hbt!]
\centering
\includegraphics[width=0.48\linewidth]{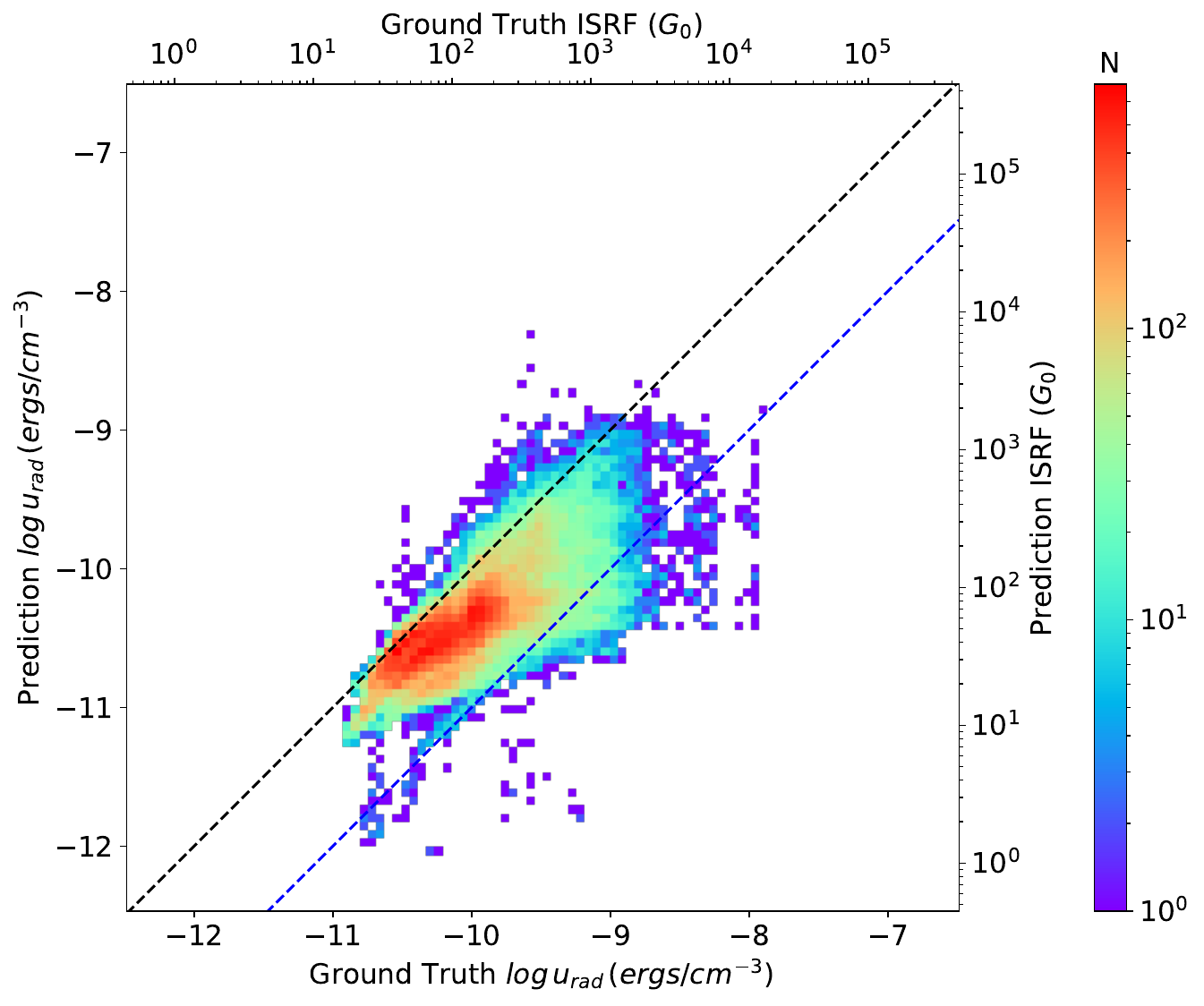}
\includegraphics[width=0.48\linewidth]{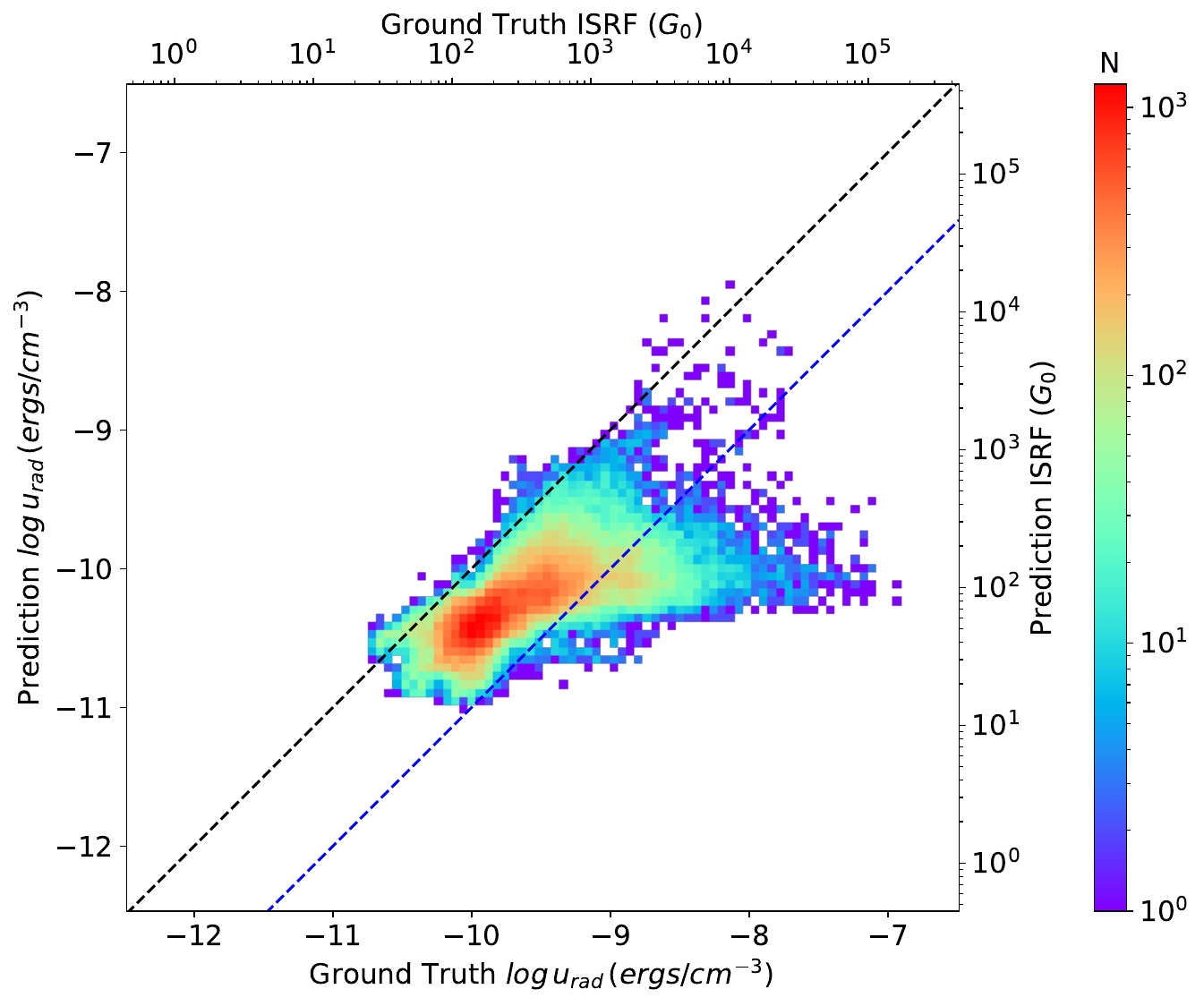}
\caption{Similar to Figure~\ref{fig.diffusion_test_hist_1}, but specifically for simulations with an ISRF of $G_{0}$ = 17 (left panel) and $G_{0}$ = 170 (right panel). The black dotted line represents the one-to-one line, which represents perfect prediction. The blue dotted line represents the ten-to-one line, indicating underestimation by a factor of 10. 
}
\label{fig.diffusion_test_hist_isrf10_100}
\end{figure*} 

Next, we assess the diffusion model's performance on new simulations characterized by significantly higher ISRF values. These simulations involve boosting the ISRF by factors of 10 and 100, resulting in ISRF intensities of $G_{0}$ = 17 and $G_{0}$ = 170, respectively. Figure~\ref{fig.diffusion_test_dust_img_isrf10_1} illustrates the radiation field predicted by the diffusion model on these simulations. Despite the large difference in the ISRF, the diffusion model still produces predictions similar to the ground truth. However, there are some discrepancies: for instance, in the third and fourth rows on the left, the presence of dotted strong radiation regions in the actual radiation field are not accurately recovered by the diffusion model. These spots are not traced by any band of the dust emission, which explains why the diffusion model may fail to recover them. 

Figure~\ref{fig.diffusion_test_hist_isrf10_100} presents a 2D histogram depicting the correlation between the diffusion model's predictions and the ground truth values of the radiation fields for simulations with ISRF intensities of $G_{0}$ = 17 and $G_{0}$ = 170. The diffusion model continues to do well, albeit returning predictions with a systematic offset. For instance, the offset between the diffusion model's predictions and the ground truth values for the 10 times higher ISRF is approximately 0.25 dex, corresponding to an underestimation factor of 1.8. Similarly, the offset for the 100 times higher ISRF is about 0.43 dex, equivalent to an underestimation factor of 2.7. Nevertheless, the relative ISRF intensity is well constrained, as the predictions and the ground truth values still exhibit a logarithmic-linear correlation, with a dispersion of 0.5. Consequently, we conclude that the diffusion model is capable of providing a reasonably accurate estimation of the radiation field even when the true field is orders of magnitude different than that of the training set. However, if a more precise estimation of the radiation field in extremely high ISRF regions is desired, it is advisable to retrain the diffusion model using an appropriate synthetic dataset.

\subsubsection{Higher Density \& Updated Heating/Cooling}

\begin{figure*}[hbt!]
\centering
\includegraphics[width=0.99\linewidth]{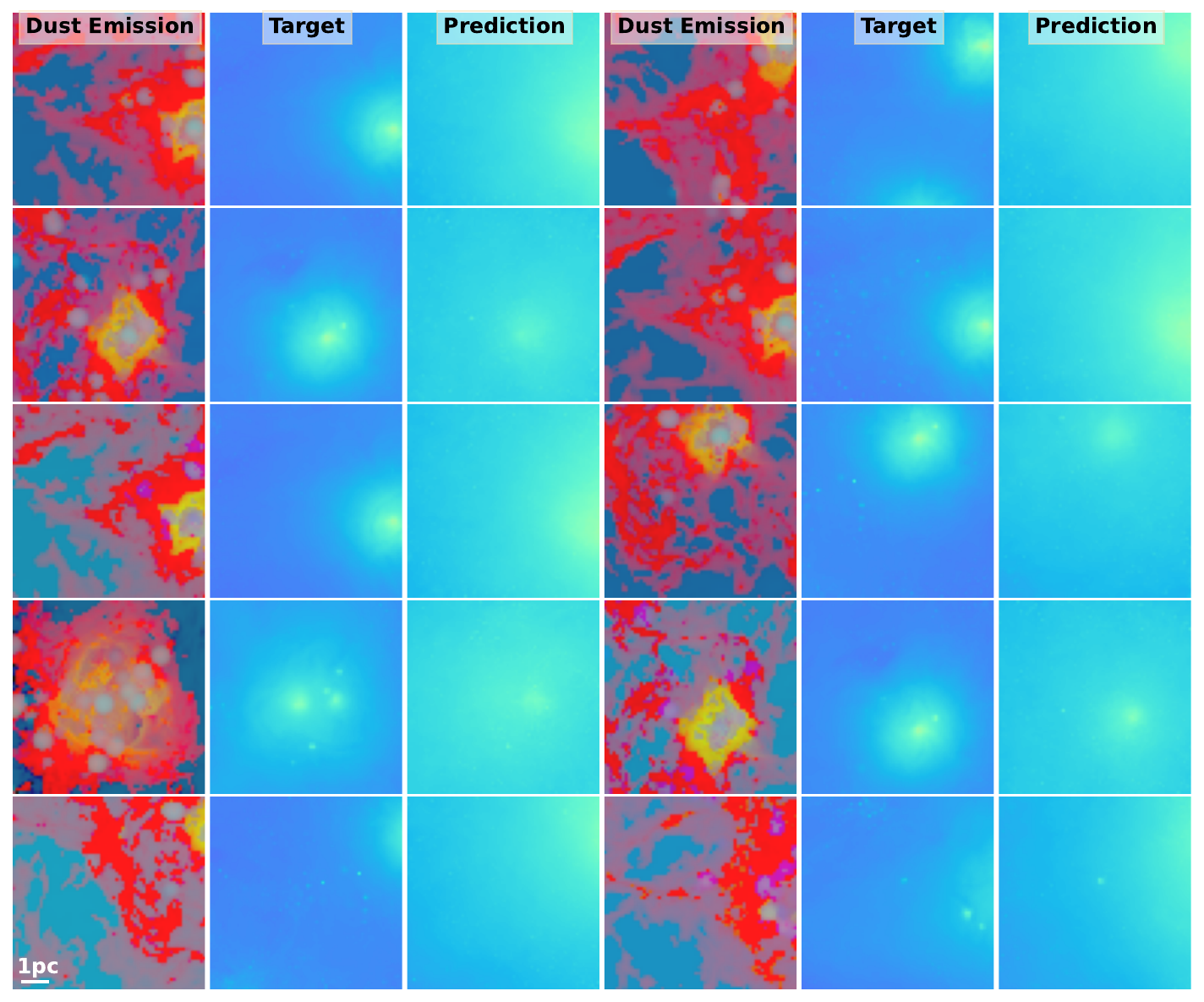}
\caption{Predicted radiation field produced by the diffusion model, alongside the associated three-band dust emission and the actual radiation field for new simulations. These simulations incorporate updated heating and cooling treatments and experience two crossing times of turbulent driving, leading to {well-developed turbulence and} increased gas density  {before self-gravity is turned on}.}
\label{fig.diffusion_test_dust_img_turbsphere_1}
\end{figure*}

\begin{figure*}[hbt!]
\centering
\includegraphics[width=0.69\linewidth]{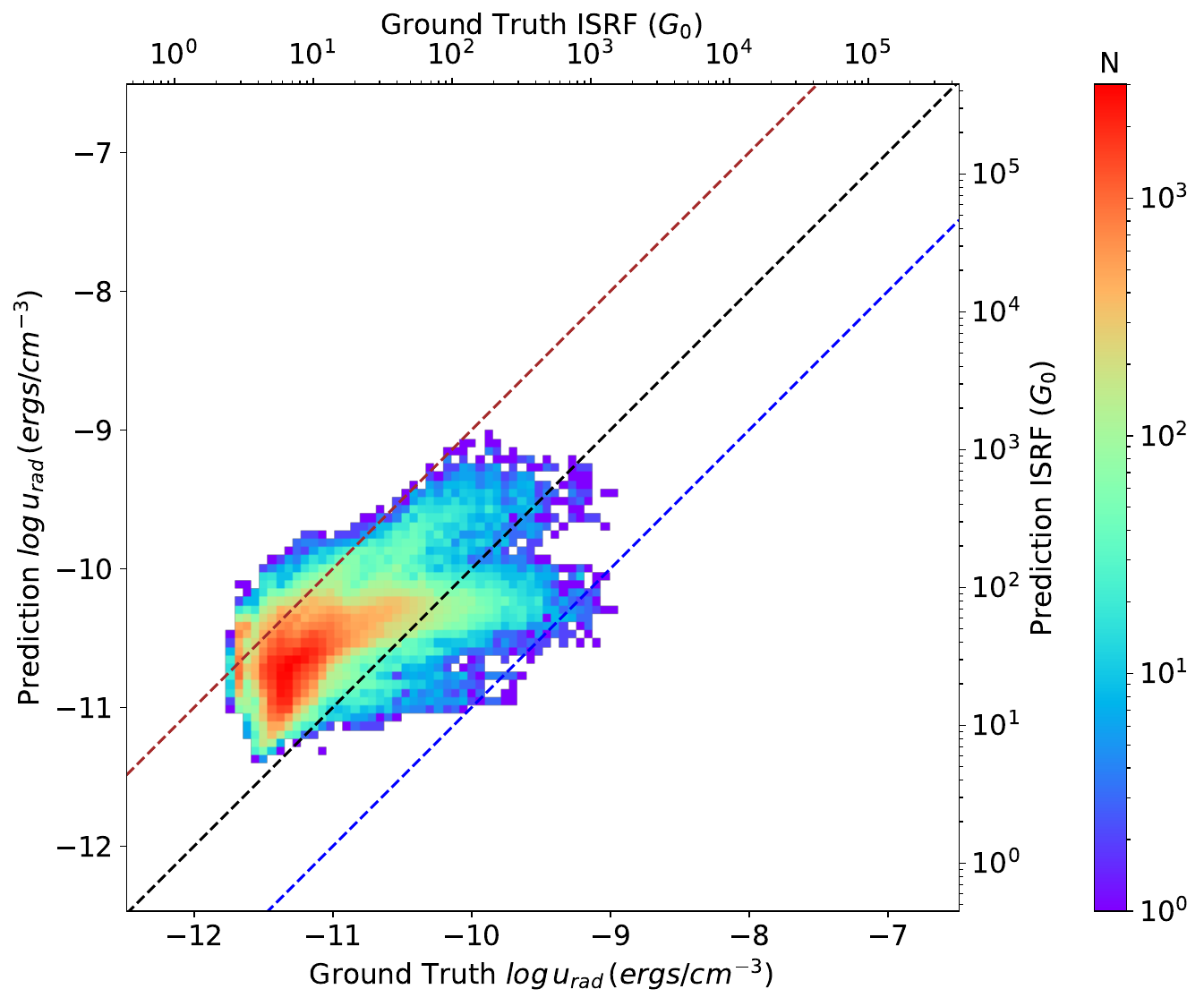}
\caption{Similar to Figure~\ref{fig.diffusion_test_hist_1}, but applied to synthetic dust images generated from new simulations with updated heating and cooling treatments, as well as two crossing times of turbulent driving. The black dotted line signifies the one-to-one line, indicating perfect predictions. The blue dotted line signifies the ten-to-one line, suggesting underestimation by a factor of 10. The brown dotted line represents the one-to-ten line, indicating overestimation by a factor of 10.}
\label{fig.diffusion_test_hist_turbsphere}
\end{figure*} 

Lastly, we evaluate the diffusion model's performance using a novel set of MHD simulations featuring different initial conditions. These simulations involve driving turbulence for two crossing times as described in \citet{2022MNRAS.510.4767L}, resulting in elevated gas density within the molecular cloud and well-developed turbulence throughout. In addition, these new simulations incorporate an updated radiative cooling and heating scheme, utilizing the cooling module shared with the FIRE-3 simulations \citep{2023MNRAS.519.3154H}. In contrast, the fiducial simulations used for training relied on a simpler fitting function based on tabulated CLOUDY results \citep{1998PASP..110..761F}, accounting for local density, temperature, and metallicity. The updated MHD simulation applied in this assessment offers a more detailed representation of heating and cooling processes, encompassing all major molecular, atomic, nebular, and continuum interactions, to better capture the thermal state of the cold ISM \citep{2023MNRAS.519.3154H}. This fresh batch of simulations allows us to assess the diffusion model's performance under extreme out-of-distribution conditions.

Figure~\ref{fig.diffusion_test_dust_img_turbsphere_1} displays the radiation field predictions generated by the diffusion model for these updated simulations. Despite significant differences in initial conditions and heating/cooling approaches, the diffusion model still generates predictions that closely resemble the ground truth, albeit with some discernible variations. Notably, the global background ISRF values predicted by the diffusion model are notably elevated compared to the ground truth.

To more precisely assess this divergence, we present a 2D histogram in Figure~\ref{fig.diffusion_test_hist_turbsphere}, illustrating the correlation between the diffusion model's predictions and the actual radiation field values for these fresh simulations. The diffusion model still performs reasonably well but exhibits a systematic offset and some variability in its predictions. It's worth noting that in these new simulations, the diffusion model appears to consistently overestimate the ISRF by approximately a factor of 3. Moreover, the dispersion in the predictions is approximately a factor of 2. 

It's possible that the updated heating and cooling methods and/or the increased density within the clumps and cores in the new simulations systematically lead to a decrease in the ISRF values. This is evident from the fact that the highest ground truth ISRF value in the new simulations is considerably smaller than that in the fiducial simulations, as observed in Figure~\ref{fig.diffusion_test_hist_1} and \ref{fig.diffusion_test_hist_dustmodel}.

Consequently, it is crucial to stress that for accurate ISRF predictions in a real molecular cloud require an appropriate training dataset reflecting the specific physical conditions of that cloud. One should exercise caution when interpreting machine learning model predictions, especially regarding their absolute values. However, the diffusion model is capable of correctly capturing relative intensity variations across various out-of-distribution datasets. This capability provides a promising avenue for assessing the relative ISRF strengths within observed molecular clouds. 

}

\subsection{Testing on MonR2}
\label{Testing on the Monoceros R2 Region}

\begin{figure*}[hbt!]
\centering
\includegraphics[width=0.98\linewidth]{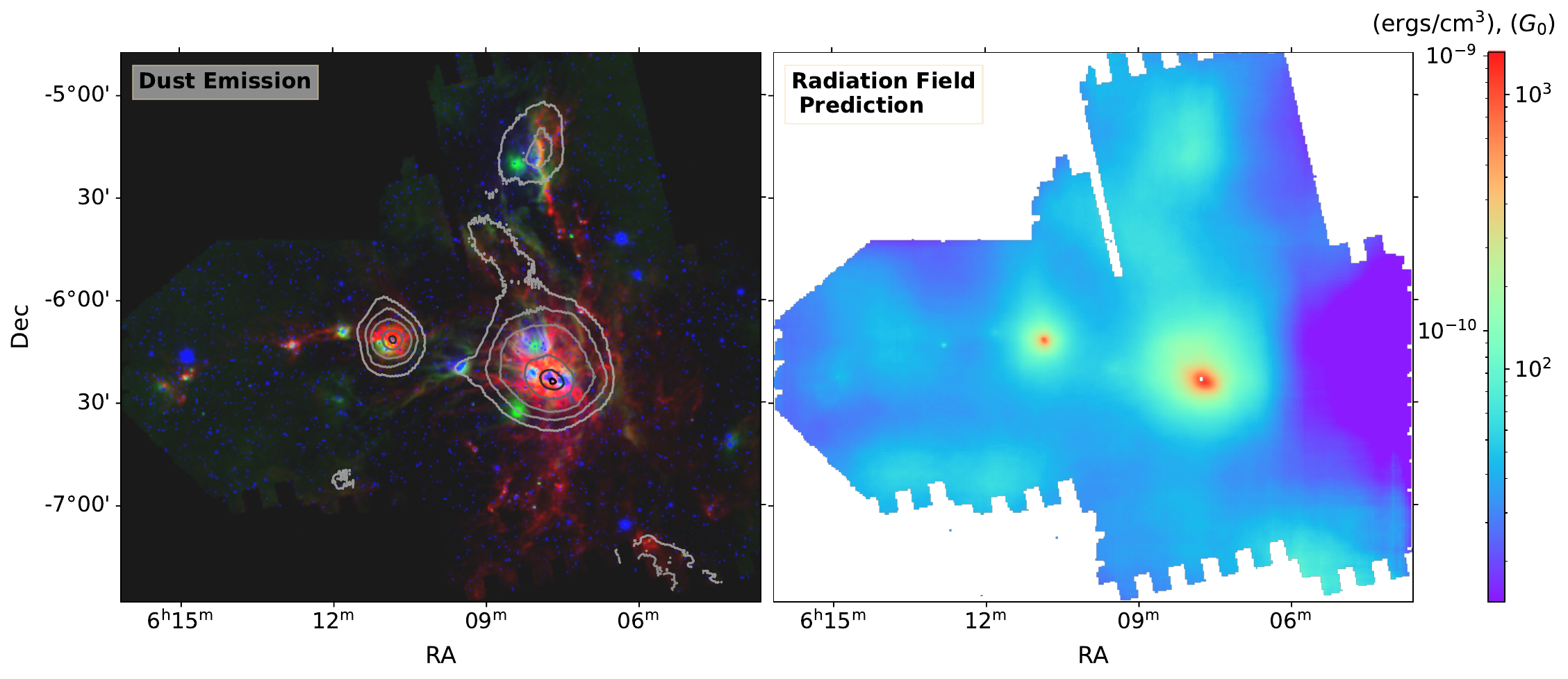}
\caption{Three-band dust emission at 4.5 \um (blue), 24 \um (green), and 250 \um (red) in MonR2 (left panel), alongside the predicted radiation field by the diffusion model (right panel). The contour lines overlaid on the dust emission maps represent the intensity of the radiation field predicted by the diffusion model. }
\label{fig.test_MonR2_pred_10}
\end{figure*}

\begin{figure*}[hbt!]
\centering
\includegraphics[width=0.98\linewidth]{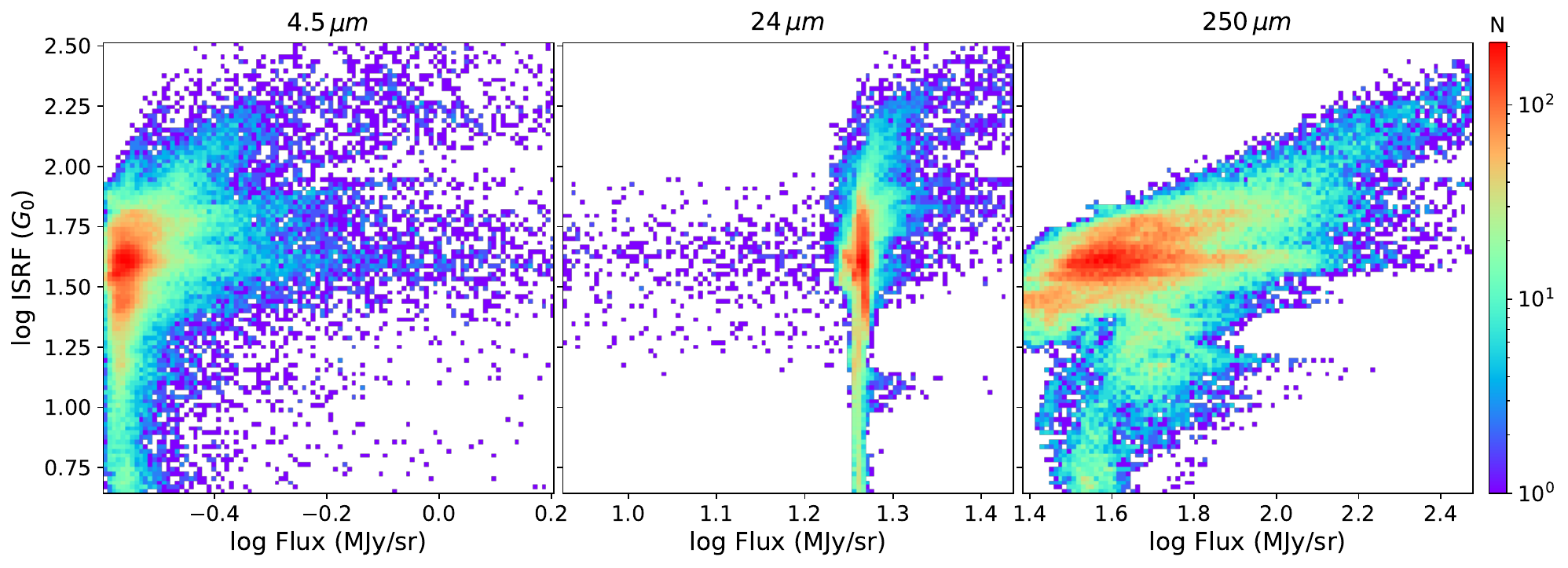}
\caption{2D histograms illustrating the correlation between the predicted ISRF and the dust emission intensity (4.5 \um\ on the left, 24 \um\ in the middle, and 250 \um\ on the right) in MonR2. }
\label{fig.correlation_isrf_infrared_monR2}
\end{figure*}

\begin{figure*}[hbt!]
\centering
\includegraphics[width=0.98\linewidth]{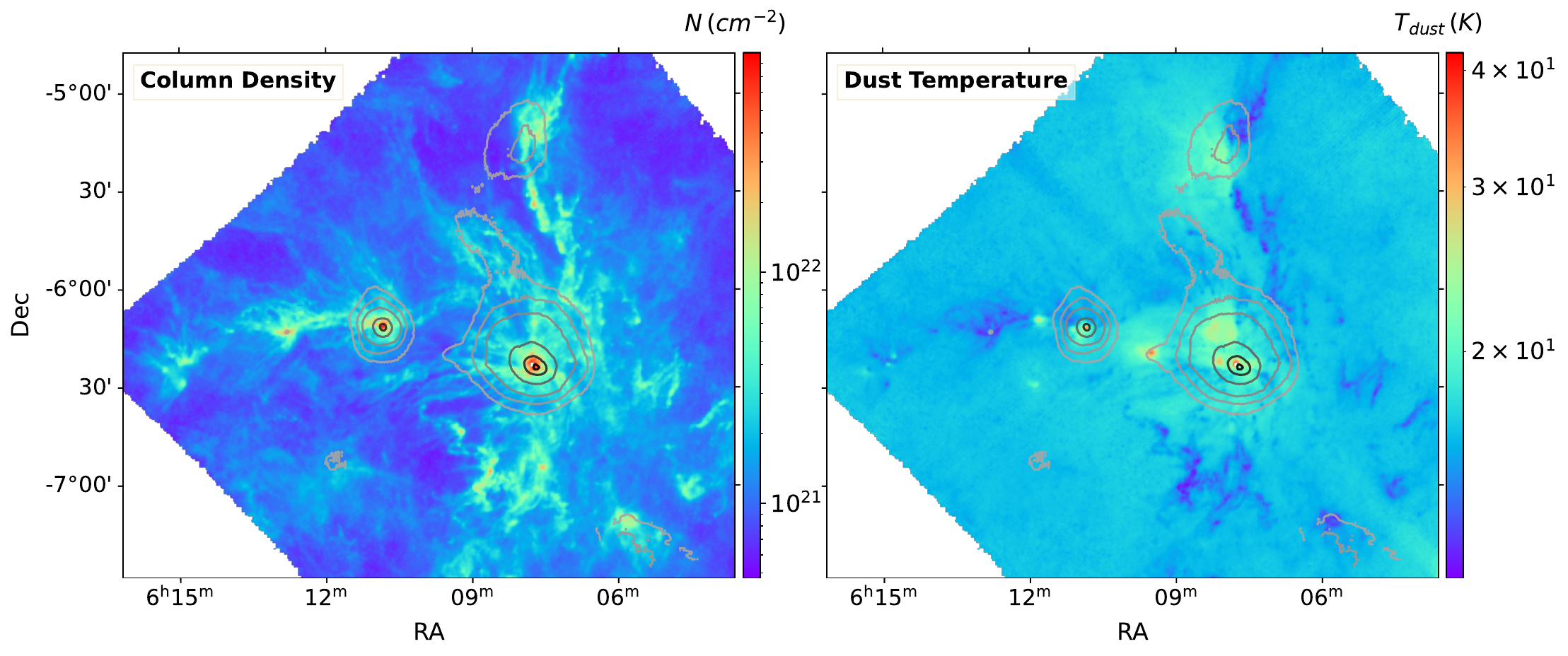}
\caption{Column density map (left) and dust temperature map (right) derived by \citet{pokhrel2016}. The contour lines represent the intensity of the radiation field predicted by the diffusion model. }
\label{fig.test_MonR2_pred_1_colden_temp}
\end{figure*}

\begin{figure*}[hbt!]
\centering
\includegraphics[width=0.98\linewidth]{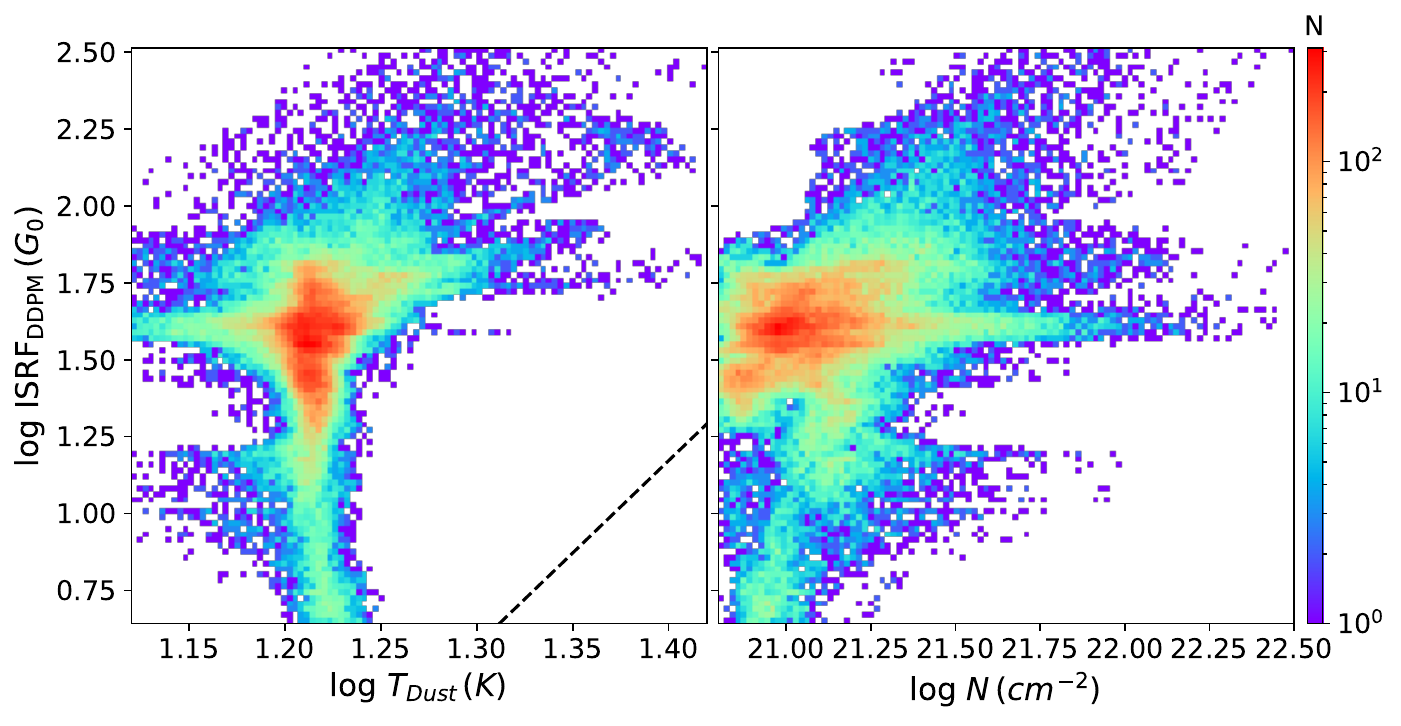}
\caption{2D histograms illustrating the correlation between the predicted ISRF and the dust temperature (left panel) and the column density (right panel) in MonR2. The black dashed line in the left panel represents the predicted relation between the ISRF and dust temperature given in Equation~\ref{equation-isrf-dust}.
%points where the ISRF$_{\rm T-Dust}$ equal ISRF$_{\rm True}$. ISRF$_{\rm T-Dust}$ is derived using the formula $(\frac{T_{\rm Dust}}{17.5\, \rm K})^{4+\beta}$ from \citet{2010A&A...518L..88B}, where the dust emissivity index $\beta$ is set to 2 in our analysis.
}
\label{fig.correlation_isrf_coldentemp_monR2}
\end{figure*}

In this section, we employ our diffusion model to analyze the actual dust observations in MonR2. The input for the diffusion model consists of three-band dust observations: 4.5 \um, 24 \um, and 250 \um. To ensure consistency in physical scales, we convolve the MonR2 dust observation to a similar physical resolution as our training data. Due to the larger size of the MonR2 dust map in terms of pixel count on both dimensions and the fixed image size requirement of the diffusion model ($64\times64$), we employ a strategy that involves cropping the large map into smaller postage stamps with dimensions of $64\times64$ and a step size of 2 pixels. After the prediction phase, these postage stamps are combined by averaging the predictions for each pixel, resulting in the reconstruction of the original large map. Figure~\ref{fig.test_MonR2_pred_10} presents the diffusion model's prediction of the radiation field. The regions of strong radiation field predominantly align with areas of intense dust emission, which aligns with our intuitive understanding that dust is most heated in regions with relatively strong radiation fields. Notably, there is some blue dotted emission that is primarily highlighted in the 4.5 \um\ band. These bright dots likely represent foreground and/or background stars that are not associated with MonR2. The diffusion model's radiation field prediction appears to successfully exclude these contaminants.

%%We find that the average projected line-of-sight radiation field intensity is significantly smaller than the fiducial value of 1 $G_{0}$. This arises 
%due to the differing definitions of the ISRF. In our work, 
%%because we average the radiation field along the line-of-sight within the molecular cloud, where radiation is attenuated due to extinction. Conversely, the traditional definition of ISRF is based on measurements taken at the surface of molecular clouds, where the dust extinction, $A_{v}$ is 0, and thus no attenuation occurs. Consequently, it is more useful to focus on the relative radiation field strength across the molecular cloud. 

Our findings demonstrate that the average projected line-of-sight ISRF in MonR2 is notably higher than the fiducial value of 1 $G_{0}$. This can be attributed to several factors. First, MonR2 is an active star-forming region where the radiation field is dominated by forming stars, leading to a significant increase in the level of ISRF. Additionally, our predicted ISRF is averaged along the line-of-sight into the molecular cloud without being extincted by dust, providing a more accurate estimation of the actual impact of radiation feedback across the cloud. %%The prediction map provides a reliable estimation of how radiation feedback impacts the MonR2 molecular cloud. 
This information is crucial for further analyses of molecular clouds, including for investigating the influence of radiation feedback on the core mass function or variation of turbulent properties.

We next perform statistical analyses to investigate the correlation between the predicted ISRF and dust emission. Figure~\ref{fig.correlation_isrf_infrared_monR2} presents 2D histograms illustrating this correlation at 4.5 \um, 24 \um, and 250 \um\ in MonR2. The dust emission at 4.5 \um\ does not exhibit a clear overall trend with the predicted ISRF. The presence of numerous blue dots, likely foreground and/or background stars unrelated to MonR2, creates a branch in the middle of the 2D histogram where the dust emission increases while the ISRF remains relatively constant. Similarly, no distinct trend is observed between the dust emission at 24 \um\ and the predicted ISRF. However, a positive correlation is evident between the dust emission at 250 \um\ and the predicted ISRF. For comparison, we investigate the correlation between the predicted ISRF and dust emission in the synthetic test data, as detailed in Appendix~\ref{Correlation of ISRF and Dust Emission in Synthetic Test Data}. We observe a similar positive trend between the dust emission at 250 \um\ and the predicted ISRF in the test data. 

Typically, longer wavelength emission, such as observed by {\it Herschel} at 160 \um, 250 \um, 350 \um, and 500 \um, is commonly used to estimate the dust column density and temperature. The dust temperature, in turn, can be used to estimate the radiation field, as shown in Equation~\ref{equation-isrf-dust}. However, as discussed in Section~\ref{Assessing the Performance of the Diffusion Model} and shown in Figure~\ref{fig.diffusion_test_hist_dust_pred}, there is only a limited correlation between the ISRF and the dust temperature in the synthetic test data. Our study extends beyond these longer wavelengths to include the analysis of shorter wavelength emission. By considering this broader wavelength range, we obtain a more accurate estimation of the stellar radiative feedback and dust emission. 

%%Typically, the dust emission at 250 \um\ is primarily influenced by dust column density rather than hot dust. However, the observed positive correlation suggests that the dust emission at 250 \um\ may also be influenced by radiation feedback, extending beyond the sole influence of column density. 

Finally, we investigate the relationship between the predicted ISRF, the column density and the dust temperature, utilizing the column density map and the dust temperature maps derived by \citet{pokhrel2016}. Figure~\ref{fig.test_MonR2_pred_1_colden_temp} showcases the column density map and the dust temperature map of MonR2. The correlation between the predicted ISRF and the column density, as well as the dust temperature, appears to be limited. While certain regions of strong ISRF coincide with areas of high column density and high dust temperature, this relationship is not consistent. Some regions with high column density and/or high dust temperature do not exhibit a strong ISRF. Figure~\ref{fig.correlation_isrf_coldentemp_monR2} displays 2D histograms illustrating the correlation in MonR2. Notably, we do not observe a clear trend between the ISRF and the dust temperature in contrast to typical modeling assumptions \citep[e.g.,][]{2010A&A...518L..88B}. Although a positive correlation can be discerned when the dust temperature exceeds 20 K (log T=1.3), it is accompanied by significant scatter. 

It is important to highlight that the black line, representing the ISRF inferred from the grey-body dust temperature, is significantly offset from the DDPM-predicted ISRF values. This discrepancy may suggest that radiation from embedded stars is highly attenudated, %% and the local ISRF is smaller than the solar neighborhood value, 
resulting in much cooler dust temperatures. Another possibility is that the dust consists of multiple temperature components, with the colder components dominating the emission in the {\it Herschel} band used to derive this temperature map. Similarly, no distinct pattern emerges between the predicted ISRF and the column density. These findings suggest that the dust emission at 250 \um\ is influenced by factors beyond just column density and dust temperature, including the presence of radiation. %%%\textbf{Isn't the Barnard relation derived assuming the solar ISRF? If so, the lower values here (even as compared to the simulation), may and suggest the local ISRF is lower, right? I added this above.  DX: I have corrected the ISRF values. The true ISRF and the DDPM predicted ISRF are significantly higher. The dust temperature inferred ISRF is smaller, primarily due to high extinction. Nevertheless, all these values are still larger than unity, i.e., log ISRF > 0. }

%SO I moved this to the end
%In the future, we plan to extend the application of the diffusion model to more archived dust observations in nearby molecular clouds, as well as to nearby galaxies. This will allow us to study the impact of radiation fields on molecular clouds and star formation in a broader context.

\section{Conclusions}
\label{Conclusions}

We produce synthetic dust observations of MHD simulations from the STARFORGE project, which incorporate various physical processes to simulate star formation and GMC evolution. Using these synthetic observations, we trained deep learning diffusion models to estimate the radiation field strength based on three-band dust emission at 4.5 \um, 24 \um, and 250 \um. We evaluated the performance of the diffusion model on both synthetic test samples and real observational data. The key findings of our study are summarized as follows:

\begin{enumerate}

%\item We performed radiative transfer simulations with various treatments of the stellar spectrum, resulting in the generation of 15,000 synthetic dust emission maps. These synthetic observations exhibit a good agreement with the SEDs observed in the Monoceros R2 region by Spitzer and Herschel telescopes.

% I think this below more than the above is what you want here, giving more detail about the dust
\item We performed radiative transfer simulations with various treatments for the spectra of stellar sources, resulting in the generation of 9,750 synthetic dust emission maps. We find that the agreement with the observed dust emission is very sensitive to the assumed dust model and show that the \citet{2023ApJ...948...55H} model for diffuse gas combined with the \citet{2017ApJS..233....1K} model for densities above $10^5$ cm$^{-3}$ provides a good representation of the MonR2 dust emission.

\item We utilized deep learning diffusion models to estimate the strength of the radiation field and assessed its performance on the test set. The diffusion model successfully reconstructed the radiation field strength at the pixel level, generally recovering the true value within 10\%.
%excellent recovery of the test set.

\item We find that the relationship between ISRF and dust temperature exhibits a high degree of scatter, such that a simple grey-body model for dust emission does not accurately predict the radiation field. 

{
\item  We evaluated the performance of the diffusion model using synthetic dust images created with different dust models than those employed in the training set. The results indicate that the diffusion model can accurately predict the ISRF, with errors within a 20\% range. This suggests that the diffusion model is not overly sensitive to our choice of dust model and can provide reliable predictions even when applied to MHD simulations with varying dust model configurations.
}

\item We assessed the performance of the diffusion model on new MHD simulations featuring an ISRF that is 10 and 100 times higher than that of the fiducial simulations. The diffusion model was still able to predict the ISRF reasonably accurately, although there was a systematic underestimation factor of 1.8 and 2.7 for the 10 and 100 times higher ISRF, respectively. 

{\item We assessed the performance of the diffusion model using new MHD simulations featuring updated heating/cooling and initial turbulent driving. The model performed satisfactorily but consistently overestimated the ISRF by a factor of 3, with a dispersion of approximately a factor of 2. This overestimation is likely due to the updated heating/cooling treatments, which systematically causes lower ISRF values in the simulations relative to simulations in the fiducial training set.

\item  The evaluation on the out-of-distribution dataset underscores the resilience of the diffusion model in predicting the relative ISRF levels within a single molecular cloud. While there are systematic offsets in the absolute ISRF values, the relative intensity of the ISRF is accurately estimated with a dispersion of up to a factor of 2. }

% As stated this isn't a finding.
% If you want you could comment on the accuracy of the typical mapping betwen derived dust temperature and ISRF (give some quantitative evaluation, based on the additional text I suggested adding above).
%\item We expect that our method exhibits robust performance across various environments, and accurately predicts different physical data when provided with appropriate training sets.

\item We employed the diffusion model to predict the radiation field in MonR2 using observed dust emission. The diffusion model successfully captures the locations of intense radiation field regions, which corresponded to areas with high dust emission. We find a positive correlation between the predicted ISRF and the dust emission at 250 \um\ with a large degree of scatter.

%SO I would remove this as a conclusion bullet. It's sort of a general issue with ML methods.
%{\item  It is vital to underscore that discrepancies in simulations featuring different initial conditions and unique physical circumstances can lead to less precise machine learning model predictions. We stress the significance of employing an appropriate training dataset that accurately reflects the particular physical conditions of the targeted molecular cloud to ensure precise ISRF predictions. It is imperative to exercise prudence when interpreting machine learning model predictions, particularly concerning their absolute values.
 %}

\end{enumerate}

%SO Revised:
{Although the model performs well overall, we stress that the test simulations, which represent out-of-distribution data, produce significantly less precise model predictions. This suggests that the uncertainties associated with the predicted ISRF represent upper limits on the accuracy of the ISRF predicted from observational data, as by definition, our training set is out-of-distribution compared to actual observations. Therefore it is important to adopt an appropriate training dataset that reflects the particular physical conditions of the targeted molecular cloud as accurately as possible to ensure the most precise ISRF prediction. }

In future work, we plan to extend the application of the diffusion model to additional archived dust observations of nearby molecular clouds, as well as to nearby galaxies. This approach will allow us to study the impact of radiation fields on molecular clouds and star formation in a broader context.

{We would like to express our gratitude to the anonymous referee for their invaluable suggestions, particularly those pertaining to the assessment of out-of-distribution data.} DX acknowledges support from the Virginia Initiative on Cosmic Origins (VICO). SSRO and RG acknowledge funding support for this work from NSF AAG grants 2107340 and 2107705. SSRO acknowledges support by NSF through CAREER award 1748571, AST-2107340 and AST-2107942,  by NASA through grants 80NSSC20K0507 and 80NSSC23K0476, and by the Oden Institute through a Moncrief Grand Challenge award. The authors acknowledge Research Computing at The University of Virginia for providing computational resources and technical support that have contributed to the results reported within this publication.

\appendix

\section{Star Categorization in Radiative Transfer}
\label{Star Categorization in Radiative Transfer}

\begin{figure*}[hbt!]
\centering
\includegraphics[width=0.98\linewidth]{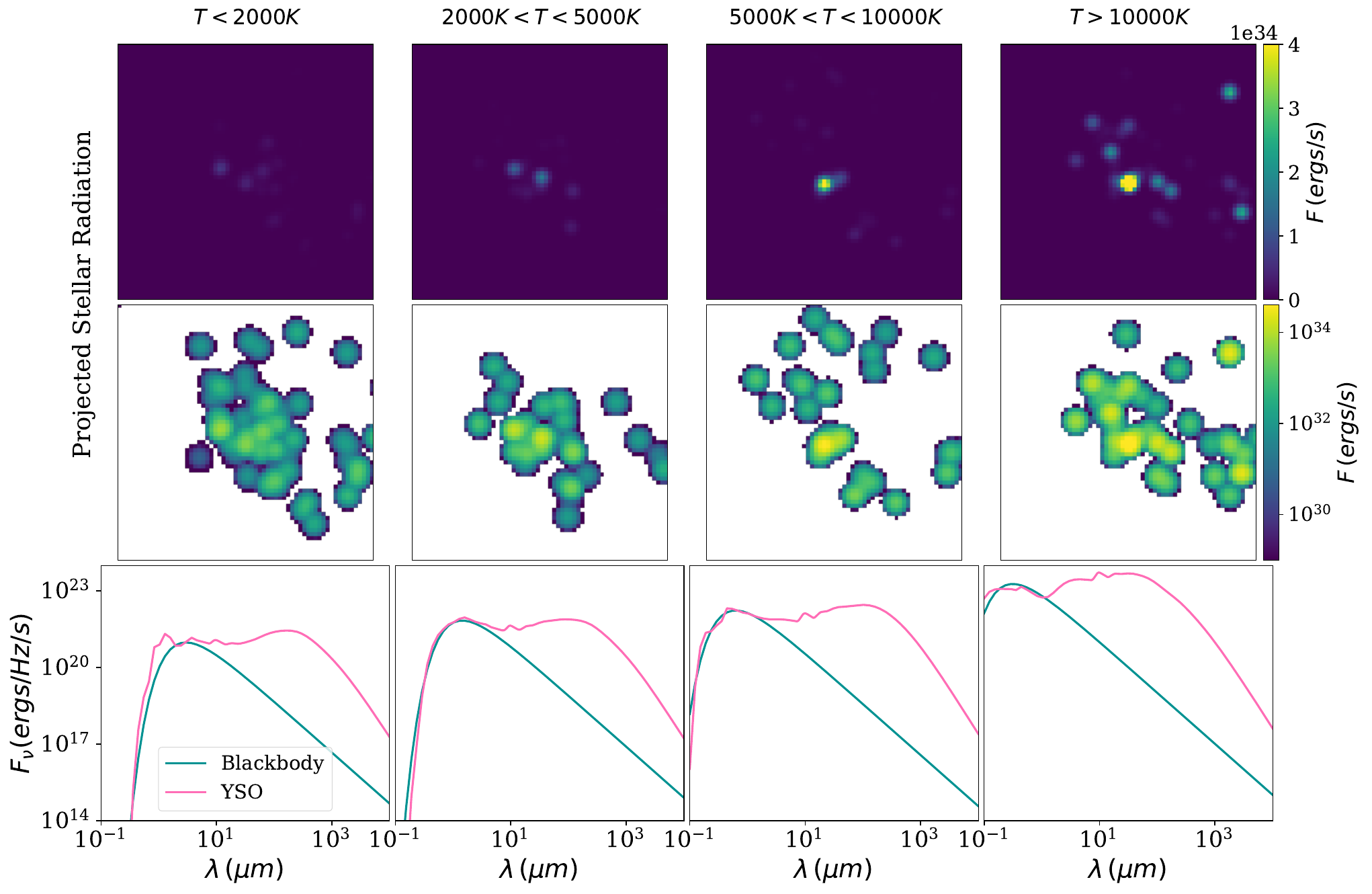}
\caption{Projected stellar radiation density and corresponding SEDs of different categories for a snapshot of simulations at 4.5 Myr without jets (the third row of Figure~\ref{fig.synthetic_dust_sed_example_1}). }
\label{fig.input_YSO_sed_stardensity_example_1}
\end{figure*} 

In this section, we provide an example of categorizing stars based on their effective temperature in radiative transfer. Due to computational limitations, it is not feasible to perform radiative transfer for each individual star in the simulation. Instead, we employ a binning approach to group stars into four categories based on their effective temperature. Figure~\ref{fig.input_YSO_sed_stardensity_example_1} illustrates the projected stellar radiation density and corresponding SEDs for a snapshot of simulations at 4.5 My without jets, as depicted in the third row of Figure~\ref{fig.synthetic_dust_sed_example_1}. The stellar field is represented by blob-like structures, deposited onto the grid with a full width at half maximum (FWHM) of 3 pixels. The SEDs for each star category are displayed, including the blackbody SED and the young stellar object (YSO) SED with disks from \citet{2017A&A...600A..11R}. Notably, the input YSO SEDs exhibit a significant excess of infrared emission due to radiative emission by an assumed circumstellar disk.%, likely attributed to dust in the disk.

\section{Exploration of Different Dust Models}
\label{Exploration of Different Dust Models}

\begin{figure*}[hbt!]
\centering
\includegraphics[width=0.98\linewidth]{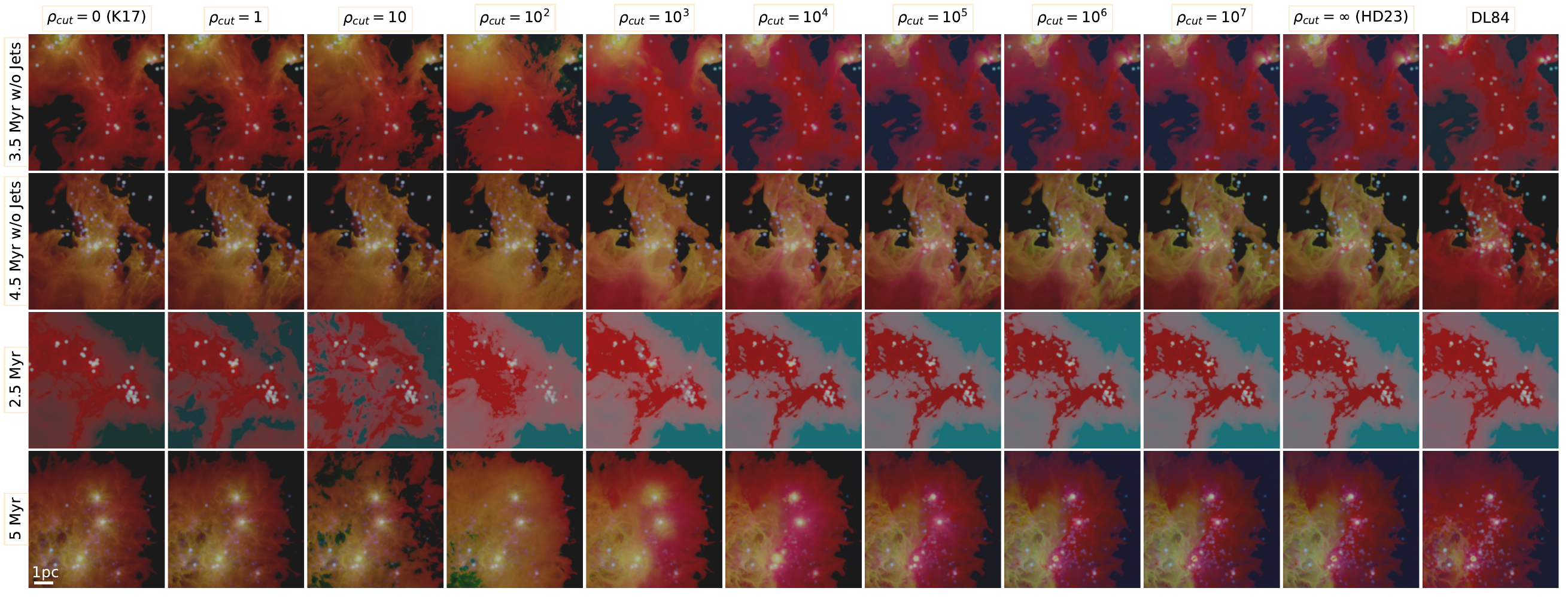}
\caption{Synthetic images at three different wavelengths (4.5 \um, 24 \um, and 250 \um) generated by employing different cutoffs ($\rho_{cut}$) on gas number densities when selecting dust models and different dust models in the radiative transfer for different simulation snapshots.}
\label{fig.synthetic_density_test_1}
\end{figure*}

\begin{figure*}[hbt!]
\centering
\includegraphics[width=0.98\linewidth]{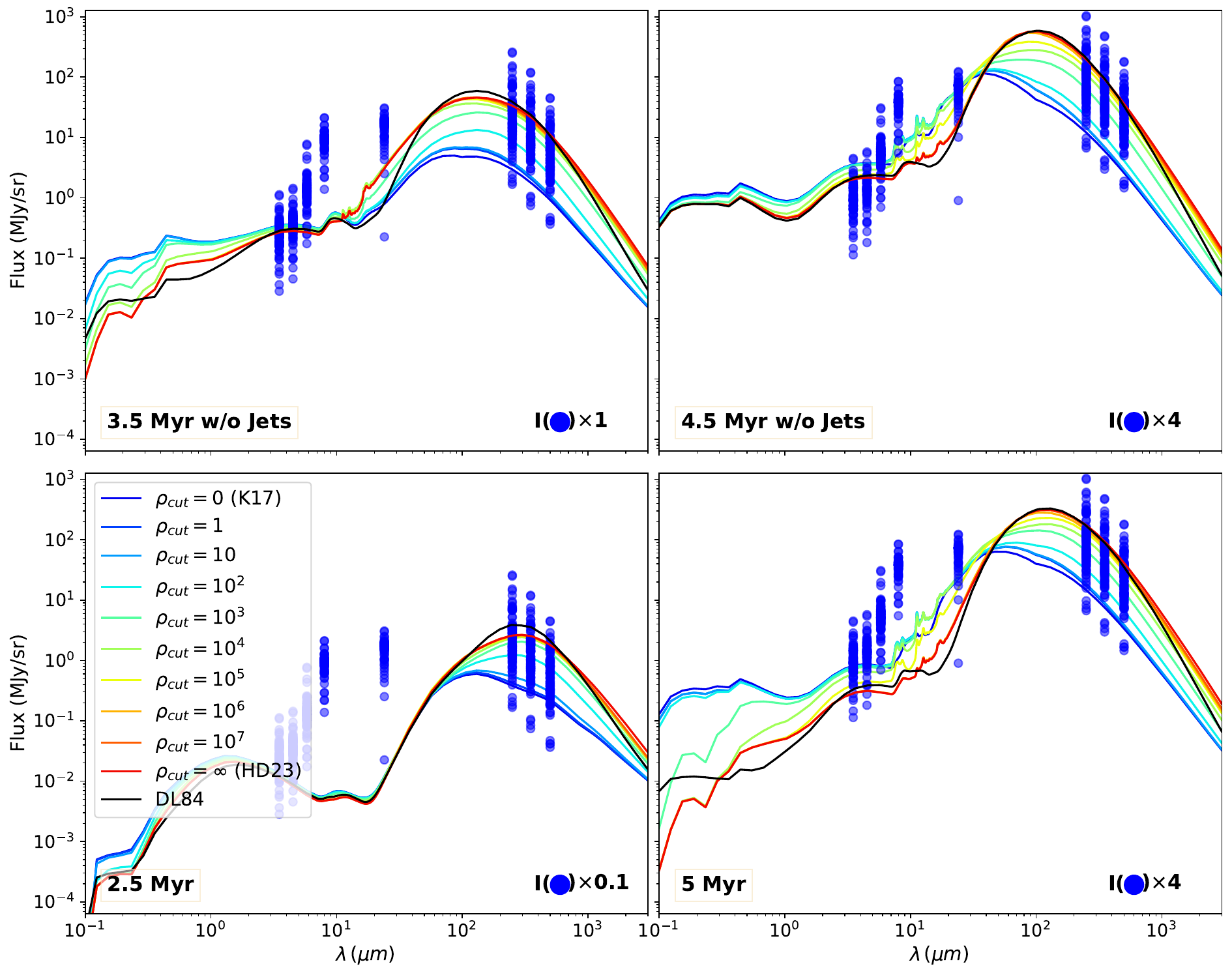}
\caption{SEDs obtained by employing different cutoffs on gas number densities when selecting dust models and different dust models in the radiative transfer for different simulation snapshots. The blue dots represent the observed SEDs in MonR2, which have been rescaled by specific factors as indicated in the lower right corner.}
\label{fig.densitycut_factor_seds_plot_1}
\end{figure*}

\begin{figure*}[hbt!]
\centering
\includegraphics[width=0.98\linewidth]{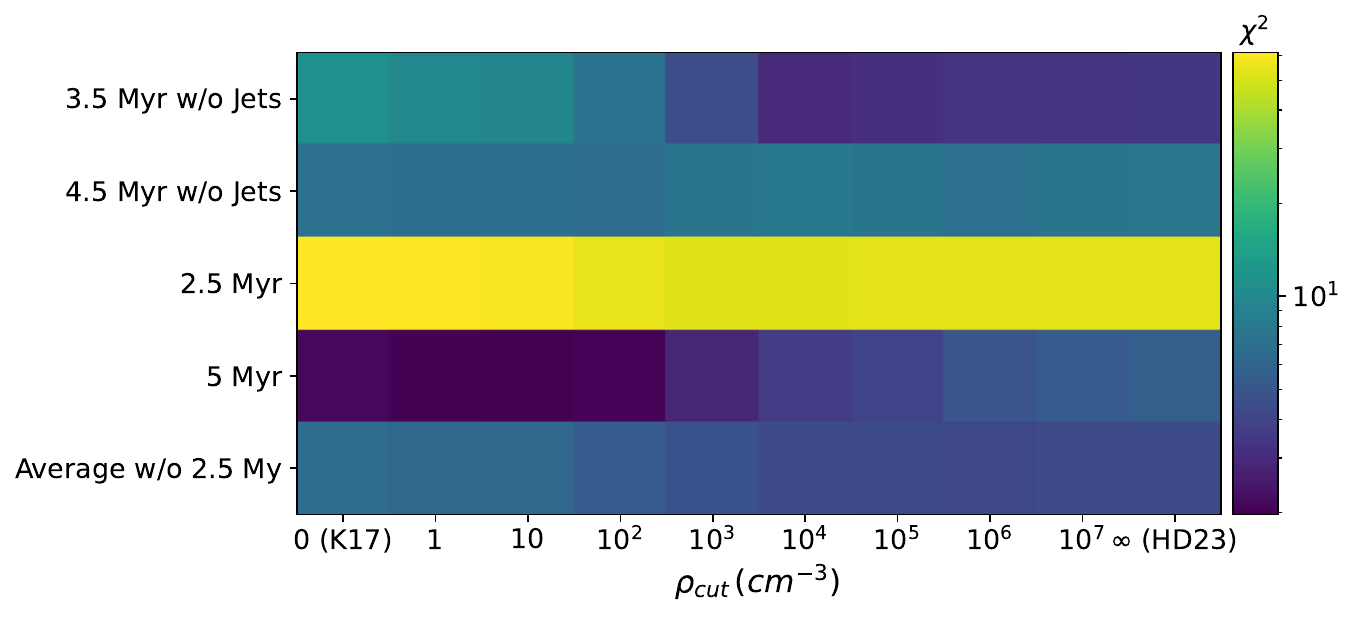}
\includegraphics[width=0.98\linewidth]{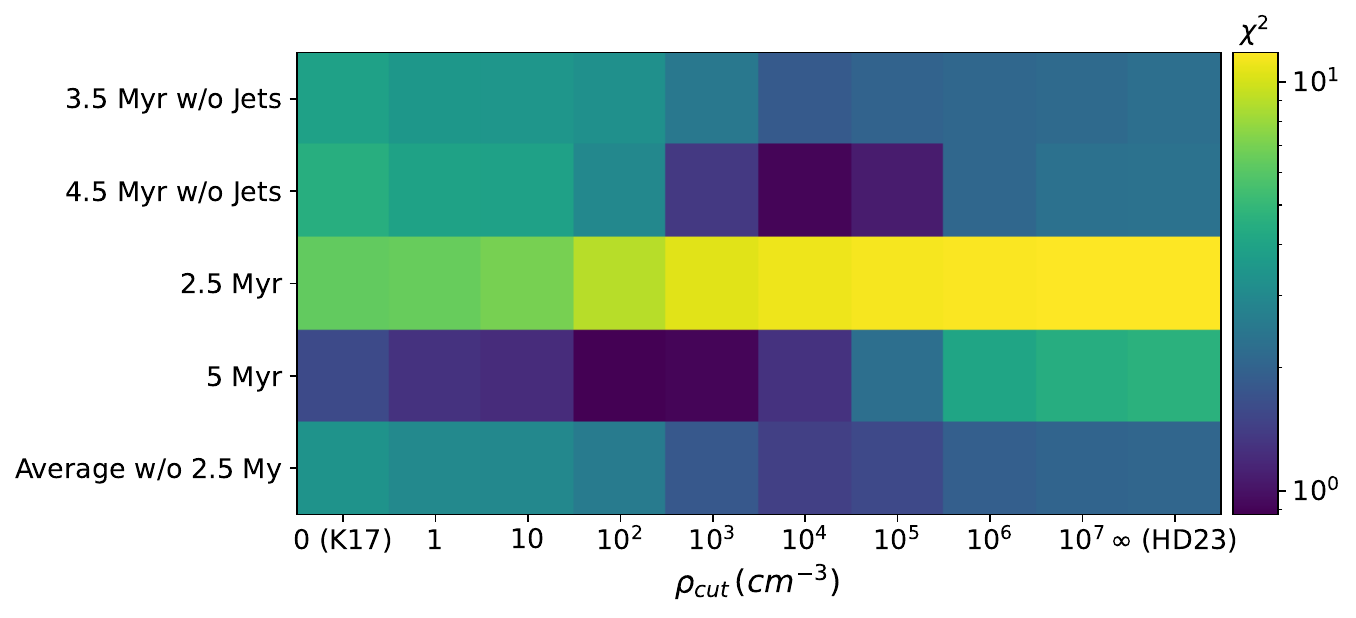}
\caption{$\chi^{2}$ values calculated between the synthetic SEDs and observed SEDs across various gas number density cutoffs during dust model selection and different evolutionary stages in various simulations. The upper panel displays the raw $\chi^{2}$ values calculated between the synthetic SEDs and the observed SEDs. In the lower panel, we compute the $\chi^{2}$ values while incorporating a free scaling parameter during the calculation. }
\label{fig.chi_square_raw_1}
\end{figure*}

In this section, we examine different cutoffs on gas number densities for selecting dust models and explore the impact of different dust models on the synthetic observations. Our primary dust model is a hybrid approach combining the K17 \citep{2017ApJS..233....1K} model for $n>10^5$ \cmc, and the HD23 \citep{2023ApJ...948...55H} model for $n<10^5$ \cmc. We vary the cutoff density, $\rho_{cut}$, from 1 \cmc~ to $10^{7}$ \cmc, as well as consider a pure K17 model (i.e., $\rho_{cut}=0$) and a pure HD23 model (i.e., $\rho_{cut}=\infty$). Additionally, we examine the performance of a traditional dust model, DL84 \citep{1984ApJ...285...89D}, on the synthetic observations.

Figure~\ref{fig.synthetic_density_test_1} and \ref{fig.densitycut_factor_seds_plot_1} depict the synthetic images in three bands (4.5 \um, 24 \um, and 250 \um) and corresponding SEDs obtained using different cutoffs on gas number densities and different dust models. The overall appearance of the synthetic images in the three bands remains similar; however, the relative intensity of the bands varies with different density cutoffs. The adoption of the K17 model results in a more yellowish color, while the HD23 model yields a more reddish color. This discrepancy indicates that the synthetic dust emission using the K17 model emits fewer long-wavelength photons compared to the HD23 model. This distinction is further evident in the SEDs, where the synthetic dust emission utilizing the K17 model exhibits weaker emission at long wavelengths but stronger emission at short wavelengths in comparison to the HD23 model. When comparing these synthetic results with the observed SEDs in MonR2, it becomes apparent that the HD23 model and the combination of the K17 model and the HD23 model better reproduce the observed emission characteristics. 

The SEDs obtained using the HD23 and DL84 models exhibit similarities, but there is a notable difference. The DL84 model fails to reproduce the PAH feature around 10\um.

{
To quantitatively evaluate the discrepancies between synthetic and observed SEDs, we provide the $\chi^{2}$ values in Figure~\ref{fig.chi_square_raw_1}. We utilize two strategies for computing these $\chi^{2}$ values. One approach directly calculates the $\chi^{2}$ value between the synthetic SEDs and the observed SEDs. The other method involves introducing a free parameter to scale the synthetic SEDs in intensity, thereby obtaining the best fit for the observed SED shape, before computing the $\chi^{2}$.  We present both for reference, since they suggest slightly different preferred dust models.
%SO Given that the data are renormalized when input to the diffusion model, the scale chi^2 is probably more representative.

It is apparent that MonR2 is not in its early evolutionary stages, as evident from the notably higher $\chi^{2}$ values for the 2.5 Myr snapshots compared to other stages. Moreover, the $\chi^{2}$ trends across different $\rho_{cut}$ values exhibit variations depending on the evolutionary stage and simulation. When averaging the $\chi^{2}$ values across simulations, excluding the early evolution at 2.5 Myr, we find that $\rho_{cut} = 10^{6}$ \cmc\ provides the best fit for the raw $\chi^{2}$, while $\rho_{cut} = 10^{4}$ \cmc\ yields the best results for the scaled $\chi^{2}$. We acknowledge that the choice of $\rho_{cut}$ may appear somewhat arbitrary since it relies on SED comparisons, which highlight the absence of a single synthetic SED that perfectly matches the observed SEDs in MonR2. By considering both methods for calculating the $\chi^{2}$ values, we opt for $\rho_{cut}=10^{5}$ \cmc. This choice can also be regarded as empirical, representing the observed distinction between diffuse and dense gas. }

\section{Correlation between the ISRF and Dust Emission in Synthetic Test Data}
\label{Correlation of ISRF and Dust Emission in Synthetic Test Data}

\begin{figure*}[hbt!]
\centering
\includegraphics[width=0.98\linewidth]{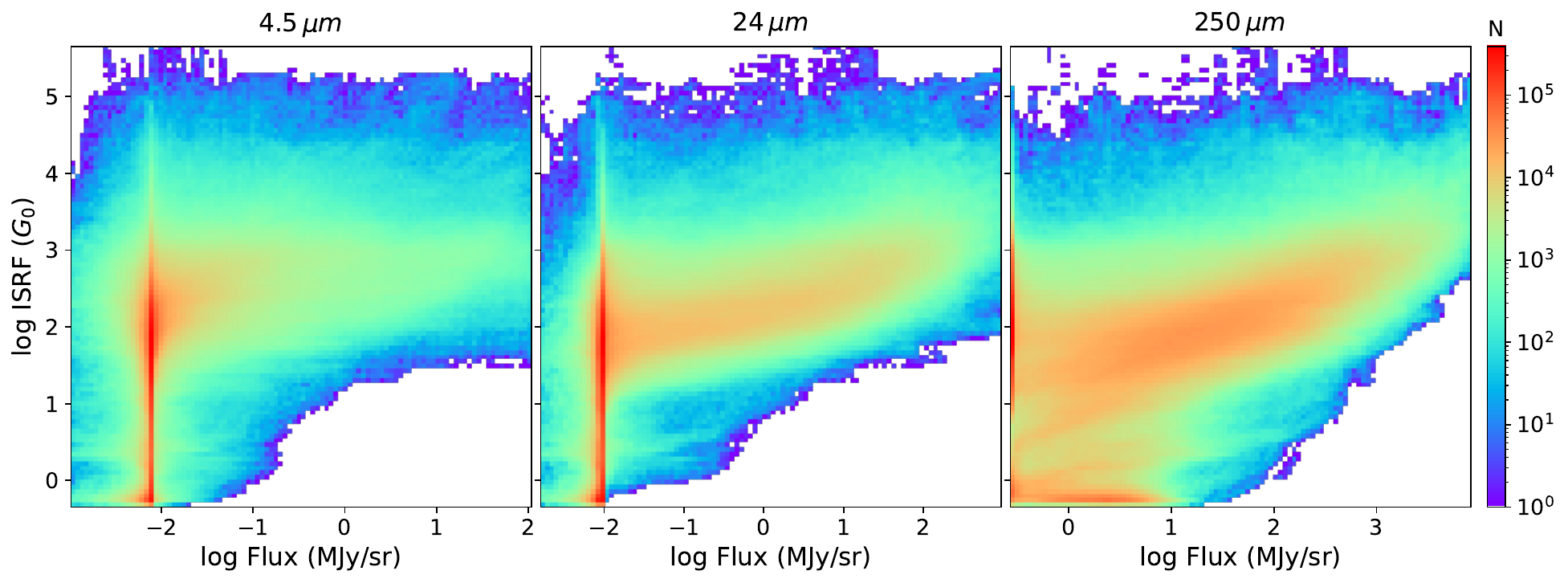}
\caption{2D histograms illustrating the correlation between the ISRF and the dust emission intensity (4.5 \um\ on the left, 24 \um\ in the middle, and 250 \um\ on the right) in the synthetic test data. }
\label{fig.correlation_isrf_infrared_synthetic}
\end{figure*}

We analyze the relationship between the ISRF and dust emission intensity in Figure~\ref{fig.correlation_isrf_infrared_synthetic}. Similar to Figure~\ref{fig.correlation_isrf_infrared_monR2}, no distinct correlation occurs between the dust emission at 4.5 \um\ and the ISRF. A weak positive correlation is apparent between the dust emission at 24 \um\ and the ISRF. A clear positive correlation occurs between the dust emission at 250 \um\ and the ISRF. This indicates that our analysis provides independent and complementary information to the typical analysis of the {\it Herschel} wavebands (160 \um, 250 \um, 350 \um, and 500 \um) used for deriving the dust temperature. Consequently, our approach presents a new metric to quantify the physical environment within molecular clouds.

\bibliographystyle{aasjournal}
\bibliography{references}

\end{CJK*}

\end{document}